\documentclass[12pt]{article}
\usepackage[margin=1in]{geometry}
\usepackage[T1]{fontenc}
\usepackage{amsmath,amssymb,amsthm,booktabs,tabularx,graphicx,setspace}
\usepackage[expansion=false]{microtype}
\usepackage[round,authoryear]{natbib}
\usepackage[hidelinks]{hyperref}
\newif\ifanon \anonfalse            

\newtheorem{theorem}{Theorem}
\newtheorem{proposition}{Proposition}
\newtheorem{lemma}{Lemma}
\newtheorem{corollary}{Corollary}
\theoremstyle{definition}
\newtheorem{assumption}{Assumption}
\newtheorem{remark}{Remark}

\newcommand{\rhoF}{\rho_F}\newcommand{\rhoH}{\rho_H}\newcommand{\rhob}{\bar\rho}
\newcommand{\Db}{\bar D}\newcommand{\kd}{k_D}\newcommand{\ke}{k_{\mathrm{eff}}}
\newcommand{\Chat}{\widehat C}\newcommand{\li}{\lambda_I}\newcommand{\lp}{\lambda_P}\newcommand{\ap}{A_P}
\newcommand{\dd}{\mathrm d}\newcommand{\Rplus}{\mathbb{R}_{+}}
\newcommand{\nChatbase}{2.21}
\newcommand{\nChatblind}{1.20}

\newcommand{\nChatfull}{3.45}

\newcommand{\nLRanblind}{0.262}

\newcommand{\nLRfair}{0.800}

\newcommand{\nPK}{1.80}
\newcommand{\nPW}{1.20}
\newcommand{\nPa}{0.50}
\newcommand{\nPaI}{1.25}
\newcommand{\nPalpha}{0.90}
\newcommand{\nPczero}{0.25}
\newcommand{\nPcm}{0.35}
\newcommand{\nPd}{0.60}
\newcommand{\nPkD}{0.02}
\newcommand{\nPkappa}{0.03}
\newcommand{\nPlam}{0.25}
\newcommand{\nPmuH}{0.50}
\newcommand{\nPmuv}{-0.28}
\newcommand{\nPpA}{0.25}
\newcommand{\nPphi}{0.60}
\newcommand{\nPq}{0.45}
\newcommand{\nPrhomax}{0.95}
\newcommand{\nPrstar}{0.40}
\newcommand{\nPsF}{0.03}
\newcommand{\nPsH}{0.70}
\newcommand{\nPsigmav}{0.75}
\newcommand{\nPt}{0.15}
\newcommand{\nPzeta}{0.80}
\newcommand{\nappBloss}{0.051}
\newcommand{\nappBrephi}{0.537}
\newcommand{\nappBreplo}{0.290}

\newcommand{\nbandhiclsblind}{2.02}

\newcommand{\nbandhiunins}{1.10}

\newcommand{\nbandloclsblind}{0.273}

\newcommand{\nbandlounins}{0.196}
\newcommand{\ncutoffDbigVbig}{0.431}
\newcommand{\ncutoffDoneVone}{0.334}

\newcommand{\ndbHdrclsblind}{-0.068}

\newcommand{\ndbHdrloading}{-0.0061}

\newcommand{\ndstarcls}{0.257}

\newcommand{\neffbase}{64}
\newcommand{\neffblind}{35}

\newcommand{\nexFdev}{-0.0494375}
\newcommand{\nexFmim}{-0.0169375}
\newcommand{\nexG}{0.1}
\newcommand{\nexHdev}{0.0169375}
\newcommand{\nexHsep}{0.0494375}
\newcommand{\nexIChi}{0.1169375}
\newcommand{\nexIClo}{0.0505625}
\newcommand{\nexbF}{0.0894375}
\newcommand{\nexbH}{0.0230625}
\newcommand{\nexbidF}{0.09}
\newcommand{\nexiota}{0.7375}
\newcommand{\nexkeff}{0.0275}
\newcommand{\nexnullbid}{0.05}
\newcommand{\nexrhs}{0.0169375}

\newcommand{\nqJ}{0.065}

\newcommand{\nqS}{0.080}
\newcommand{\nqswitch}{0.800}
\newcommand{\nqpeakclsdhigh}{0.155}
\newcommand{\nqpeakclsdlow}{0.820}
\newcommand{\nqpeakclsdmid}{0.255}

\newcommand{\nqpeakexpdhigh}{0.333}
\newcommand{\nqpeakexpdlow}{0.720}
\newcommand{\nqpeakexpdmid}{0.427}

\newcommand{\nrentclsblind}{49}
\newcommand{\nrentclsfull}{97}

\newcommand{\nrentexpblind}{92}

\newcommand{\nrhoF}{0.173}
\newcommand{\nrhoH}{0.838}

\newcommand{\nrhocutoffDoneVone}{0.579}

\newcommand{\nrobn}{1000}

\newcommand{\nshiftcls}{0.068}
\newcommand{\nshiftcrosssub}{0.062}
\newcommand{\nshiftexp}{0.0061}

\newcommand{\nshiftloading}{0.0061}

\newcommand{\nvtopclsblind}{2.02}
\newcommand{\nvtopclsblindbigK}{52.55}
\newcommand{\nvtopclsfull}{2.87}

\newcommand{\nvtopexpblindbigK}{5.67}

\newcommand{\nwtpcls}{0.112}

\newcommand{\nxbcls}{0.494}
\newcommand{\nxbexp}{0.162}
\newcommand{\nzobsH}{0.896}

\title{\bf Insuring the Fallback:\\ Capital, Monitoring, and the Certification of\\ Preserved Human Capability under Improving AI}
\ifanon
\author{}
\date{}
\else
\author{Andreas Bauer\thanks{Aegis Compliance and Strategies O\"U, Tallinn, Estonia.
Correspondence: \texttt{a.bauer@science.us.org}. ORCID 0000-0002-6539-9353. This paper
builds on four companion papers, \citet{bauer2026lastsignal,bauer2026fallbacksignal,bauer2026window,bauer2026threeceilings};
Table~\ref{tab:prov} states the provenance of every model component. Replication code and every
reported number are archived at doi:10.6084/m9.figshare.33193704. All errors are my own.}}
\date{September 2026}
\fi

\begin{document}
\maketitle

\begin{abstract}
\noindent
When generative AI makes the deliverable uninformative, a professional-services provider can still
certify the preserved human capability to catch the machine's errors, through a liability pledge
whose expected cost falls in that capability. The pledge is credible only up to what can be
collected, and that ceiling is set by an underwriter, which bears part of the pledge and audits the
insured. The range of client stakes over which one certificate separates has a width bounded by the
provider's own capital plus the audited share of the underwriter's capacity, so unmonitored capacity
adds nothing to it. Blind capital instead relocates that range upward, through a cross-subsidy that
exists only under class rating. Monitoring converts capital into width, removes the cross-subsidy,
raises the return to preserving skill, and, because audit information leaks, makes the certificate
redundant beyond an interior precision.
\end{abstract}

\medskip
\noindent\textbf{Keywords:} liability insurance; costly signalling; monitoring; class rating; credence goods; human--AI complementarity.\\
\noindent\textbf{JEL:} D82, D86, G22, K13, L15.

\newpage
\section{Introduction}\label{sec:intro}

A radiology group, an audit practice and a translation agency have each concluded that the artifact
no longer proves anything. Once the machine writes as well as the expert, a fluent report is evidence
about the machine, not about whoever signed it \citep{bauer2026lastsignal}. What remains informative
is the residual: what happens in the fraction of cases where the system is wrong, and whether anyone
on the provider's side has stayed practised enough to notice. That capability is built by keeping
people engaged, it decays when they are not, and it can be certified by a liability pledge, because a
provider who has preserved it expects to pay damages less often
\citep{singh2026fallback,bauer2026fallbacksignal}.

The pledge has a ceiling. A promise larger than what can be collected is not a promise, and the
existing treatment of that ceiling is a single exogenous solvency parameter. In the markets this
literature is about, the ceiling is not a parameter. It is a price quoted by a professional-indemnity
underwriter, in a market that as of 2026 is being rewritten in both directions at once: affirmative
AI liability towers appeared during 2025--26 with per-organisation aggregate limits in the eight
figures, while general liability forms began carrying generative-AI exclusion endorsements from
January 2026 \citep{ailiabilitymarket2026,verisk2026}. The party that supplies the credibility of the
competence signal is a strategic actor, and it is an actor with a distinctive property: it can look at
the thing being signalled. An underwriter that audits training logs, exception-handling drills and
staffing rosters observes the insured's engagement policy far more directly than any client ever
will.

This paper asks what that does to the signal.

\subsection*{The two objects}

Let the provider retain a share $r$ of the pledge and cede $1-r$ to an underwriter that loads the
premium at rate $\li$ and prices the ceded part off an audit of precision $q$: with probability $q$
the audit reveals the insured's type, otherwise the premium is set at a tariff reference. Write
$t=\varphi p_A$ for the probability that an AI failure is unrescued-and-verified against the fringe,
and $x_T=1-\rho_T$ for the unrescued share of type $T$. The expected cost of a pledge $D$ to a
provider of type $T$ is
\begin{equation*}
C_T(D)=tD\big[\underbrace{\big(r+(1+\li)(1-r)q\big)}_{\iota}\,x_T
+\underbrace{(1+\li)(1-r)(1-q)}_{\lp}\,x_b\big],
\end{equation*}
where $x_b$ is the unrescued share at the tariff reference. We call $\iota$ the \emph{informational
retention}: it is the coefficient on the type-dependent part of the cost, hence the coefficient on the
difference between a competent provider's cost and a mimic's. We call $\lp$ the \emph{pooled load}:
cost that the provider pays and that carries no information about it. Retention and monitoring are
substitutes in producing $\iota$; blindness and cession are complements in producing $\lp$.

Three things follow, and the first two correct the natural reading of the 2026 market.

\paragraph{(i) Capital without monitoring does not extend the reach of a certificate; it relocates
it.} The set of client stakes over which \emph{one} certificate separates a competent provider from a
mimic is an interval, and its width is proportional to the \emph{effective informational capacity}
$\Chat=\iota\Db$, where
$\Db(r)=\min\{W/r,K/(1-r)\}$ is what the provider's own capital $W$ and the underwriter's capacity
$K$ jointly support. Theorem~\ref{thm:capacity} shows that $\max_r\Chat(r,q)=W+(1+\li)qK$ at
$r^\ast=W/(W+K)$: with $q=0$ the maximum is $W$ exactly, for every $K$. Buying a larger tower from an
underwriter that does not look does not add one euro to the range of engagements any single
certificate can discriminate. What it does (Theorem~\ref{thm:frontier}) is shift that range upward,
by an amount that under experience rating
is the loading alone and vanishes at actuarially fair prices, and under class rating is the
cross-subsidy the competent provider pays for being pooled with the fringe. In the scenario, doubling
$K$ from \nPK\ to twice that raises the largest certifiable ticket under a blind class-rated carrier
while leaving the width at $t\Delta\rho W$; raising $K$ to 100 raises it to \nvtopclsblindbigK\ under
class rating and to \nvtopexpblindbigK\ under experience rating, against \nvtopclsfull\ under a full
audit at $K=\nPK$.

\paragraph{(ii) Monitoring is what converts capital into credibility, and it does so at a linear
rate.} The marginal informational product of capacity is $(1+\li)q$: zero for a blind carrier, linear
thereafter. The retention that attains the maximum, $r^\ast=W/(W+K)$, does not depend on $q$: the
design of the optimal programme is invariant to how well the underwriter looks; monitoring changes
its value, not its structure. Under class rating monitoring has a second effect: it removes the
cross-subsidy, so that the competent provider's share of the certification rent rises from
\nrentclsblind\% under a blind carrier to \nrentclsfull\% under a full audit.

\paragraph{(iii) Cover erodes the incentive to preserve the fallback through two channels, and
monitoring closes one of them.} Ceding exposure raises the competent provider's certificate cost
through the loading, always, and through the cross-subsidy, under class rating only. The return to
becoming the competent type therefore falls in the cession share, and the second channel is removed
by monitoring (Proposition~\ref{prop:invest}). This is the classical moral-hazard reading of
insurance, located in a specific channel and with a specific remedy.

\subsection*{The third audience}

The underwriter is a third audience for the competence signal, after clients and mobile workers, but
it is not additive, because it is the only audience that can inspect the type directly. Whatever it
learns can leak: professional-indemnity terms are disclosed in tenders, brokers publish ratings,
coverage schedules appear in procurement packs. Let $d$ be the share of the underwriter's information
that reaches clients. Then monitoring has two opposed effects on the provider's own signal: it
sharpens and cheapens the pledge, and it makes the pledge redundant, because the public shadow of the
audit already separates the provider from the fringe. The result is a \emph{certification window}:
the share of client stakes over which a certificate is available is hump-shaped in $q$ whenever
disclosure is high, with the peak moving left as disclosure rises---from $q=\nqpeakclsdlow$ at
$d=0.35$ to $q=\nqpeakclsdhigh$ at $d=0.95$ in the scenario (Proposition~\ref{prop:window}).

\subsection*{What is done and what is not}

The equilibrium object is deliberately coarse. An institution offers a binary menu---a null contract
and one certificate---and the retention, audit precision and tariff are public before the provider's
type is drawn. Under those restrictions the entire separating equilibrium reduces to two inequalities
(Theorem~\ref{thm:binary}), and every capacity, monitoring and disclosure result is a statement about
those two inequalities. We show in Appendix~B why the natural alternative---a continuum of types
separated by a continuous pledge schedule, the construction of \citet{bauer2026fallbacksignal}---does not deliver an equilibrium once the null contract and the
client's expected recovery are priced: the bottom type prefers not to participate, and once the
client pays for the indemnity the marginal pledge is fairly priced to the provider and stops
separating. Coarseness is not a simplification here; it is the reason the instrument works.
Section~\ref{sec:coarse} shows that the binary model extends to a continuum of types as a cutoff
equilibrium, so nothing in the results depends on there being two types.

Two objections are worked through rather than left to a limitations paragraph: that the
underwriter's audit responds to the retention the provider chooses (Proposition~\ref{prop:global}
gives the exact global condition under which $r^\ast$ survives), and that the underwriter screens at
the point of sale (screening enters the redundancy channel, not the retention). We also separate
what a market with censored verification records can and cannot establish: the recorded success rate
overstates the rescue probability by a known factor and can be corrected at rate $N^{-1/2}$
(Proposition~\ref{prop:statistics}), but a corrected record is not incentive compatibility, and we do
not claim that it is.

\subsection*{Contribution}

To the literature on insurance as a quality signal---of which \citet{zhang2022returninsurance} is the
closest instance---we add a signalled type that is produced, depreciating and directly observable by
the insurer, and we show that the insurer's monitoring, not its capital, is what determines whether
the instrument separates. To the insurance-economics literature on monitoring and private regulation
\citep{benshahar2012outsourcing,holzapfel2024ubi} we add an externality: the underwriter's audit
decision determines whether a third party, the client, can tell one provider from another, and the
underwriter does not capture that value. To the economics of AI and work
\citep{acemoglu2026collapse,lovett2026commons,caosun2026augmentation} we add the observation that the
private market which finances skill preservation is bounded not by willingness to pay but by
underwriting practice---and by which of two standard rating conventions the underwriter uses. To the
credence-goods tradition \citep{darby1973free,dulleck2006doctors,dulleck2011economics} we add a case
in which the disciplining instrument is itself supplied by a third party with superior information.

Section~\ref{sec:lit} places the paper. Section~\ref{sec:model} sets up the model.
Sections~\ref{sec:eqm}--\ref{sec:coarse} contain the results, Section~\ref{sec:records} the
statistical companion, Section~\ref{sec:data} an empirical design, Section~\ref{sec:disc} the
discussion. Proofs are in Appendix~A; Appendix~B explains why the continuum construction is not
inherited; the online appendix reports the scenario values, robustness and simulation details.

\section{Related literature and the gap}\label{sec:lit}

Five streams meet here. We take them in turn and state what each does not do.

\paragraph{Insurance as a signal.} That buying cover can reveal private information is established.
\citet{thakor1982exploration} shows that third-party information production by a guarantor can
sustain a signalling equilibrium in debt insurance; \citet{gracerebello1993financing} show that higher
corporate coverage can signal favourable private information to investors.
\citet{zhang2022returninsurance} is the sharpest customer-facing statement: a retailer's adoption of
return insurance signals quality to consumers, the insurer sets premiums strategically, and the third
party produces a double marginalisation that strengthens the instrument. Three features separate that
model from this one. First, their signalled type is an exogenous property of a product; ours is a stock
of human capability produced by an engagement decision and eroding when that decision is not renewed.
Second, their insurer prices off a belief; ours can \emph{audit}, and the whole content of
Theorems~\ref{thm:capacity} and~\ref{thm:frontier} is what happens when the third party's information
precision is a choice variable. Third, their instrument is bought or not bought; here what is chosen
is how much of a pledge to retain. The corporate-insurance lineage
(\citealp{holmstrom1979moral} on informativeness; \citealp{rothschild1976equilibrium} on screening in
insurance markets) supplies the machinery but not the object.

\paragraph{Insurance as monitoring and as private regulation.} \citet{benshahar2012outsourcing} argue
that insurers reduce moral hazard through underwriting, pricing and loss control, and that regulation
can usefully be outsourced to them. \citet{holzapfel2024ubi} give the closest formal analogue to our
window: a monitoring technology is adopted only if it is sufficiently accurate, which explains the
sluggish diffusion of usage-based cover. What this literature studies is the effect of monitoring on
the \emph{insured's care}. What it does not study is the effect of monitoring on a \emph{third
party's inference}. Our client never contracts with the underwriter and yet the underwriter's
monitoring decision determines whether the client can tell one provider from another at all. That
externality is the paper's subject, and it has a rating-convention dimension the care literature does
not need: under class rating a blind audit pools the competent provider with the fringe, and the
cross-subsidy it pays is the channel through which capital relocates the market.

\paragraph{What kind of monitoring this is.} In the costly-state-verification tradition of
\citet{townsend1979optimal}, the uninformed party pays to observe a state ex post, contingent on the
informed party's report, and the optimal contract has a verification region
\citep{galehellwig1985incentive,mookherjee1989optimal}. Our underwriter chooses the \emph{precision of
a signal ex ante} and prices off the resulting posterior; nothing is triggered by a report. The model
belongs to the informativeness tradition of \citet{holmstrom1979moral} and \citet{shavell1979risk},
and $\iota$ is an informativeness statistic. The insurance side of that principle has its own canon:
\citet{arrow1963uncertainty} founds insurance-plus-moral-hazard in a medical credence-good market,
\citet{pauly1968economics} names the incentive cost of coverage, and \citet{shavell1979moral}
characterises the partial-coverage optimum that is the ancestor of every retention here. What
\emph{is} costly state verification in this paper is $\varphi$: whether a failure can be proved ex
post, inherited from \citet{bauer2026lastsignal} and a property of the legal environment.

\paragraph{Coverage limits and rating conventions.} Ceilings on coverage are not an artefact of our
setup. \citet{huberman1983indemnity} derive upper limits on indemnity from the limited liability that
bankruptcy confers; \citet{bergesio2025optimal} show that under limited liability full insurance is
not optimal even at fair prices and that the optimum is a capped deductible. A separate objection runs
the other way: since large limits worsen incentives, carriers ration capacity ex ante
\citep{winter2006liability,parra2025moralhazard}. That strengthens Corollary~\ref{cor:exclusions}: a
carrier that rations limits because it cannot observe care is a carrier with $q$ near zero, which is
the regime in which the limits would not have widened the market anyway. The distinction between
class rating and experience rating that organises Section~\ref{sec:capacity} is textbook
\citep{dionne2013handbook}; its role here is that it determines whether the tariff reference is the
competent type or the pool, and hence whether blind capital carries a cross-subsidy.

\paragraph{Certification, disclosure, and signal redundancy.} The redundancy channel of
Section~\ref{sec:monitoring} has its home in the certification literature surveyed by
\citet{dranovejin2010quality}. \citet{lizzeri1999information} shows that a monopoly certifier reveals
only whether quality clears a minimal standard; \citet{daughety2008communicating} study disclosure and
a costly signal jointly; \citet{grossman1981informational} distinguishes the warranty argument from
inference about the warranted product. That literature studies what a certifier chooses to reveal.
It does not study what the revelation does to a \emph{separate costly signal the certified party was
already posting}, which is the object here: the underwriter's disclosure does not merely inform the
client, it destroys the instrument the provider was using to inform the client itself.

\paragraph{The fallback problem and deskilling.} \citet{bainbridge1983ironies} named the irony; the
modern versions are quantitative. \citet{budzyn2025deskilling} document adenoma-detection decay among
experienced endoscopists after routine AI exposure; \citet{bastani2025guardrails} show that unguarded
assistance impairs acquisition in novices, a different object we keep separate. \citet{caosun2026augmentation}
give the firm-side micro-foundation for why the stock erodes, and \citet{lovett2026commons} state the
circularity: effective oversight depends on the expertise that adoption undermines.
\citet{acemoglu2026collapse} place the mechanism at the level of aggregate knowledge; their collapse
requires a \emph{joint} condition---elastic human effort and agentic accuracy above a threshold---and
is the macro counterpart of Corollary~\ref{cor:betterai}. \citet{singh2026fallback} turn it into a
labour problem with an endogenous, appropriable stock; on the operations side
\citet{devericourt2026delegating}, \citet{boyaci2024human} and \citet{vaccaro2024combinations} study
the human--machine interface, but in none of it does the human--AI system face a buyer inferring
quality from a contract. None of it prices the stock, and none asks who finances the instrument that
would.

\paragraph{Division of labour with the companions.} \citet{bauer2026lastsignal} establishes that
expected client recovery enters the price of a pledged service, an accounting convention we adopt and
which turns out to matter (Appendix~B). \citet{bauer2026fallbacksignal} supplies the rescue technology
and a continuous-type benchmark; \citet{bauer2026window} the participation, enforcement and capped
signal-set logic; \citet{bauer2026threeceilings} the distinction between provability, financial
capacity and legal caps. There the credible ceiling is exogenous and provability moves; here
provability is a scalar and the ceiling is the endogenous object. The two sets of comparative statics
compose: every boundary in this paper scales in $\varphi$, so the companion's state-dependent
provability transmits one-for-one. Table~\ref{tab:position} locates the paper.

\begin{table}[t]
\centering\small
\caption{Position of the paper relative to the closest work.}
\label{tab:position}
\begin{tabularx}{\textwidth}{@{}lXXX@{}}
\toprule
& \citet{zhang2022returninsurance} & \citet{bauer2026fallbacksignal} & This paper \\
\midrule
Signalled type & exogenous product quality & endogenous fallback skill & endogenous fallback skill \\
Third party & insurer prices the premium & absent (ceiling exogenous) & underwriter chooses cession, audit, rating convention \\
Insurer information & belief only & --- & audit of precision $q$; class or experience reference \\
Instrument & adopt / not adopt & continuous pledge schedule & binary certificate menu plus retention $r$ \\
Credibility ceiling & not modelled & exogenous & $\Db(r)=\min\{W/r,K/(1-r)\}$ \\
Central result & insurance separates; double marginalisation & liability certifies preserved skill & capital widens the market only through monitoring; blind capital relocates it through the cross-subsidy; certification window in $(q,d)$ \\
\bottomrule
\end{tabularx}
\end{table}

\paragraph{The gap.} The pledge that certifies preserved competence has a ceiling; the ceiling is
supplied by a party that can observe what is being certified; nobody has asked what that party's
information and rating convention do to the signal. The answer inverts the natural policy reading of
the 2026 insurance market: a deepening of affirmative AI liability capacity, unaccompanied by
underwriting that inspects human-in-the-loop practice, does nothing for the market's ability to
\emph{discriminate} preserved competence, and under class rating it moves that market away from the
engagements where the discrimination is cheapest to buy.

\clearpage
\section{Model}\label{sec:model}

\subsection{Primitives}

\paragraph{Technology and the fallback.} Jobs are handled by an AI system that fails with probability
$p_A\in(0,1)$, independently across jobs. Conditional on a failure, a provider of type $T\in\{F,H\}$
rescues the job with probability $\rho_T$, $0\le\rhoF<\rhoH<1$. Write $x_T=1-\rho_T$ for the
unrescued share and $\Delta\rho=\rhoH-\rhoF=x_F-x_H$ for the rescue gap. Types have strictly positive
prior probabilities, $\mu_H\in(0,1)$ for $H$. Section~\ref{sec:coverage} interprets the type as a
skill stock produced by engagement, using the rescue technology
$\rho(s)=\rho_{\max}(s/\alpha)^a$ of \citet{bauer2026fallbacksignal} and the learning--erosion
transition of \citet{singh2026fallback}; Section~\ref{sec:coarse} replaces the two types by a
continuum. All parties are risk neutral.

\paragraph{Verification and pursuit.} An unrescued failure is attributable and provable with
probability $\varphi\in(0,1]$. Write $t=\varphi p_A$. Verification precedes the client's decision to
pursue payment. Pursuit costs the client $c_0>0$ and recovers the pledge $D$ with certainty once
verification has occurred, so the client pursues iff $D\ge c_0$ and nets $D-c_0$. Clients cannot
affect outcomes, prevent a rescue or fabricate verification; this excludes the double moral hazard of
\citet{cooperross1985warranties}.

\paragraph{Capital and the pledge.} The provider ring-fences capital $W>0$ and the underwriter
capacity $K>0$ for collectible losses, net of all fees and premiums. For an observable ticket $v>0$
an institution fixes a policy $(r,q)$ and a binary menu $\{0,D\}$ before the provider observes its
type. The positive certificate costs $\kd>0$ to administer and must satisfy
\begin{equation}
c_0\le D\le\Db(r)\equiv\min\{W/r,\ K/(1-r)\},
\label{eq:cap}
\end{equation}
with $\Db(0)=K$ and $\Db(1)=W$. The null contract costs nothing. Only these two contracts trade;
providers post no further terms, retentions or private prices.

\paragraph{The tariff and its two references.} The provider retains a share $r\in[0,1]$ of the
pledge; the underwriter loads the premium at $\li\ge0$, $a_I\equiv1+\li$, and audits the insured's
engagement record with precision $q\in[0,1]$: with probability $q$ the audit reveals the type, and
the premium is the loaded true expected loss; otherwise the premium is the loaded expected loss at a
\emph{tariff reference} $\rhob\in[\rhoF,\rhoH]$, $x_b=1-\rhob$. The expected premium for true type
$T$ is
\begin{equation}
P_T=a_I(1-r)\,tD\,\{q\,x_T+(1-q)\,x_b\}.
\label{eq:premium}
\end{equation}
The reference is the unaudited underwriter's best estimate of its insured, and it is pinned by the
composition of the book the underwriter prices:
\begin{itemize}\itemsep1pt
\item under \emph{experience rating}, the policy is written on the class of providers who post
certificates; on the separating path that class contains only $H$, so $x_b=x_H$;
\item under \emph{class rating}, the policy is written before certification on the whole
professional class, so $x_b=1-(\rhoF+\mu_H\Delta\rho)$, the unrescued share at the prior mean.
\end{itemize}
Both are Bayesian; they differ in what the underwriter's book contains when it prices. Real
professional-indemnity underwriting is class-rated with experience adjustments, so the class-rated
case is the empirically natural one and the experience-rated case is the informational benchmark.
Every result is stated for general $x_b\in[x_H,x_F]$ and interpreted at the two anchors. Neither the
audit outcome nor the realised premium is a public signal.

\paragraph{Client value includes indemnity.} Following \citet{bauer2026lastsignal}, expected recovery
enters the client's valuation. Let $\zeta\in(0,1]$ scale the client's willingness to pay for quality
and $u\in(0,1]$ the residual demand for the provider's own signal ($u=1$ in the baseline;
Section~\ref{sec:monitoring} sets $u=1-dq$). Write $\chi_v=\zeta v p_A u$ and $G(v)=\chi_v\Delta\rho
=gv$, $g=\zeta p_Au\Delta\rho$. A competitive buyer who believes the rescue probability to be
$\widehat\rho$ values the certificate, relative to fringe service, at
\begin{equation}
\chi_v(\widehat\rho-\rhoF)+t(D-c_0)(1-\widehat\rho),
\label{eq:bid}
\end{equation}
the second term being expected recovery net of pursuit expense. The null contract carries no
indemnity; its bid is $B_0(\widehat\rho)=\chi_v(\widehat\rho-\rhoF)$, which is zero at belief $F$ and
$\mu_HG$ at the prior.

\subsection{Timing}

\begin{enumerate}\itemsep1pt
\item The institution announces $(r,q)$, the menu $\{0,D\}$ and the tariff reference.
\item Nature draws the type, privately observed by the provider.
\item The provider chooses the null contract or the certificate.
\item Buyers observe the choice and pay the competitive bid \eqref{eq:bid}.
\item The underwriter audits; the premium is \eqref{eq:premium} in expectation.
\item Jobs execute; unrescued verified failures trigger payment $D$, split $r$/$1-r$; the client
pursues iff $D\ge c_0$.
\end{enumerate}
The equilibrium concept is perfect Bayesian equilibrium. We claim existence of a separating
equilibrium under exact conditions and characterise when it coexists with null pooling; we do not
claim uniqueness or a refinement.

\subsection{Assumptions}

\begin{assumption}[Regularity]\label{as:reg}
$0\le\rhoF<\rhoH<1$; $\varphi\in(0,1]$, $p_A\in(0,1)$; $\li\ge0$; $W,K>0$; $c_0,\kd>0$;
$\mu_H\in(0,1)$; tickets are drawn from a distribution $F$ with a continuous density on $\Rplus$.
\end{assumption}

\begin{assumption}[Commitment and competitive pricing]\label{as:commit}
A posted pledge is enforceable and cannot be renegotiated after a verified failure. The tariff is
\eqref{eq:premium} with a reference determined by the underwriter's book as described above; the
underwriter earns the loading and no informational rent on the certified path. Coverage is a
proportional quota share.
\end{assumption}

\begin{assumption}[No ex-ante screening]\label{as:noscreen}
The underwriter's only information technology is the audit of precision $q$, private to the insurance
contract except for a disclosed share $d$ (Section~\ref{sec:monitoring}). Access to capacity is not
rationed on type.
\end{assumption}

Assumption~\ref{as:noscreen} is the one a reader of the corporate-insurance literature will want to
attack, and rightly; Section~\ref{sec:objections} drops it.

\subsection{What is inherited and what is new}

\begin{table}[t]
\centering\small
\caption{Provenance of model components.}
\label{tab:prov}
\begin{tabularx}{\textwidth}{@{}Xl@{}}
\toprule
Component & Source \\
\midrule
Verifiability $\varphi$, pledge $D$, expected recovery in the client's price & \citet{bauer2026lastsignal} \\
Rescue technology $\rho(s)$, client credence stage, stakes $\chi$ & \citet{bauer2026fallbacksignal} \\
Skill transition, learning and erosion rates & \citet{singh2026fallback} \\
Null option, participation floor, capped signal set & \citet{bauer2026window} \\
Distinction between financial and legal ceilings & \citet{bauer2026threeceilings} \\
Retention $r$, audit $q$, loading $\li$, tariff \eqref{eq:premium}, two rating references & new \\
Informational retention $\iota$, pooled load $\lp$, effective capacity $\Chat=\iota\Db$ & new \\
Binary-menu equilibrium, width and frontier theorems, null-pooling condition & new \\
Certification window, monitoring market, investment channels, cutoff extension & new \\
\bottomrule
\end{tabularx}
\end{table}

\begin{table}[t]
\centering\small
\caption{Notation and scenario values. Scenario values are dimensionless; those marked
``inherited'' are taken from the companions for comparability, those marked ``new'' are set here.
None is an estimate.}
\label{tab:notation}
\begin{tabularx}{\textwidth}{@{}lXll@{}}
\toprule
Symbol & Meaning & Scenario & Source \\
\midrule
$p_A$, $\varphi$, $t$ & AI failure, verification, $t=\varphi p_A$ & \nPpA, \nPphi, \nPt & inherited \\
$\rho_{\max},a,\alpha$ & rescue technology $\rho(s)=\rho_{\max}(s/\alpha)^a$ & \nPrhomax, \nPa, \nPalpha & inherited \\
$s_F$; $\rhoF$ & fringe skill; its rescue probability & \nPsF; \nrhoF & inherited \\
$s_H$; $\rhoH$ & competent skill; its rescue probability & \nPsH; \nrhoH & new \\
$c_0$, $\kd$ & pursuit cost floor, certificate administration cost & \nPczero, \nPkD & inherited \\
$\zeta$ & client's quality-value share & \nPzeta & inherited \\
$\mu_v,\sigma_v$ & log-normal ticket distribution (illustrative) & \nPmuv, \nPsigmav & new \\
$W$, $K$ & provider capital, underwriter capacity & \nPW, \nPK & new \\
$\li$, $a_I$ & loading & \nPlam, \nPaI & new \\
$r^\ast$ & capital-balancing retention $W/(W+K)$ & \nPrstar & new \\
$q$, $d$ & audit precision, disclosure & \nPq, \nPd & new \\
$\mu_H$ & prior probability of the competent type & \nPmuH & new \\
$x_b$ & tariff reference: experience-rated / class-rated & \nxbexp\ / \nxbcls & new \\
$c_m$, $\kappa$ & audit cost scale, cost of becoming the competent type & \nPcm, \nPkappa & new \\
\bottomrule
\end{tabularx}
\end{table}

Tables~\ref{tab:prov} and~\ref{tab:notation} state the provenance of every component and fix
notation. Symbols for verification, the pledge and AI capability retain the meanings of the published
\citet{bauer2026lastsignal}; $m$ is reserved for the legal-cap multiple of the companions and is not
monitoring.

\clearpage
\section{The certificate equilibrium}\label{sec:eqm}

\subsection{Informational retention and pooled load}

\begin{lemma}[Cost decomposition]\label{lem:decomp}
Under Assumptions~\ref{as:reg}--\ref{as:commit}, the expected cost of a pledge $D$ to a provider of
type $T$, retained damages plus premium, is
\begin{equation}
C_T(D)=tD\,\big[\iota\,x_T+\ap\big],\qquad
\iota=r+a_I(1-r)q,\quad \lp=a_I(1-r)(1-q),\quad \ap=\lp x_b.
\label{eq:cost}
\end{equation}
\end{lemma}

Three features of \eqref{eq:cost} are worth stating before any result is derived from them.

\emph{Retention and monitoring are substitutes in $\iota$, and each alone suffices.} $\iota>0$
whenever $r>0$ or $q>0$. A provider that retains everything needs no audit; one that cedes everything
needs a perfect one.

\emph{The loading is a signalling subsidy.} At $q=1$, $\iota=1+\li(1-r)>1$ whenever $\li>0$ and
$r<1$. A fully audited, loaded premium is \emph{more} type-sensitive per unit of pledge than
self-retention, because a mimic pays $a_I$ times its true excess loss. Insurance loading, normally a
deadweight, here amplifies the difference between a competent provider's cost and a mimic's.

\emph{The pooled load is deadweight for signalling.} $\lp$ raises the cost of every pledge by the
same amount irrespective of type. The share of the pledge's cost that carries information,
$\iota/(\iota+\lp)$, equals one under a perfect audit and $r/(r+a_I(1-r))$ under a blind carrier:
\neffbase\% against \neffblind\% at $r^\ast$ in the scenario. Figure~\ref{fig:retention}(a) maps
$\iota$.

\subsection{Single crossing}

\begin{proposition}[Single crossing under insurance]\label{prop:sc}
$C_F(D)-C_H(D)=tD\,\iota\,\Delta\rho$. The difference between a mimic's and a competent provider's
cost of the same pledge is strictly increasing in $D$ if and only if $\iota>0$. At $r=q=0$ the two
types' costs coincide for every $D$, and no certificate can be strictly incentive compatible.
\end{proposition}

The economics is one sentence. Damages are owed only in states where the AI failed and the human did
not rescue; a competent provider is in that state less often; but the provider only \emph{feels} that
saving to the extent that it either pays the damages itself or is charged for them by someone who has
looked. What insurance does, absent monitoring, is convert a type-contingent cost into a pooled
one---which is exactly what a signal cannot survive.

\subsection{The binary-menu equilibrium}

On a separating path the buyer's bid for the certificate is \eqref{eq:bid} at $\widehat\rho=\rhoH$,
so the provider receives $G(v)+t(D-c_0)x_H$, and the null bid is zero. Define the net certificate-cost
coefficients after the on-path indemnity receipt,
\begin{align}
b_H&=t\{(\iota-1)x_H+\ap\}=t(1-r)\{\li x_H+a_I(1-q)(x_b-x_H)\}\ \ge0,\label{eq:bH}\\
b_F&=t\{\iota x_F+\ap-x_H\}=b_H+t\iota\Delta\rho,\label{eq:bF}
\end{align}
and $\ke=\kd+tx_Hc_0>0$. The competent type's certificate payoff is $G-\ke-b_HD$; a mimic's is
$G-\ke-b_FD$; both earn zero at the null contract under separating beliefs.

\begin{theorem}[Binary-menu separation]\label{thm:binary}
Fix the policy and a feasible pledge $D$. A separating perfect Bayesian equilibrium in which $F$
chooses the null contract and $H$ the certificate exists if and only if
\begin{equation}
\boxed{\ \ke+b_HD\ \le\ G(v)\ \le\ \ke+b_FD.\ }
\label{eq:icir}
\end{equation}
With both inequalities strict, both provider choices are strict.
\end{theorem}

The left inequality is the competent provider's participation constraint against the null contract;
the right is the mimic's deterrence constraint. Because the action set contains exactly these two
choices, the two inequalities are the complete equilibrium condition: there is no off-path pledge
whose belief has to be specified, and no appeal to a refinement.

\begin{corollary}[Tickets and menus]\label{cor:tickets}
Suppose $u>0$, $\iota>0$ and $\Db(r)\ge c_0$. The set of tickets for which \emph{some} separating
menu exists is exactly
\begin{equation}
v\in\Big[\frac{\ke+b_Hc_0}{g},\ \frac{\ke+b_F\Db}{g}\Big]\equiv[v_{\rm lo},v_{\rm hi}],
\label{eq:tickets}
\end{equation}
and for such a ticket the set of feasible separating pledges is
\begin{equation}
\Big[\max\{c_0,(G-\ke)/b_F\},\ \min\{\Db,(G-\ke)/b_H\}\Big],
\label{eq:menus}
\end{equation}
the last quotient omitted when $b_H=0$. If $\Db(r)<c_0$ no certificate is feasible.
\end{corollary}

\begin{figure}[t]
\centering\includegraphics[width=.9\textwidth]{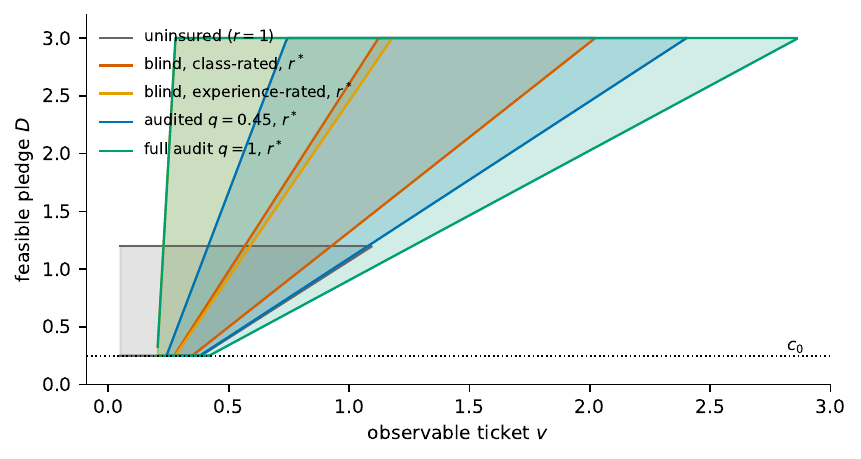}
\caption{\textbf{The feasible certificate set.} For each observable ticket $v$, the interval of
pledges $D$ that support separation (Corollary~\ref{cor:tickets}), under five configurations at the
scenario values; $d=0$. The set is bounded below by the larger of the pursuit floor $c_0$ and the
mimic's deterrence pledge, and above by the smaller of the capital ceiling and the competent
provider's participation pledge. Blind class-rated cover shifts the set toward larger tickets;
monitoring widens it.}
\label{fig:menus}
\end{figure}

Figure~\ref{fig:menus} plots the feasible pledge interval over tickets for five insurance
configurations. It is the replacement for a continuous pledge schedule, and it is a set, not a
function: for a given ticket many pledges separate, and the smallest deterring pledge frequently
leaves the mimic indifferent. Two observations organise everything that follows. The interval in
$G$-space has width $(b_F-b_H)D=t\iota\Delta\rho D$, which does not depend on $x_b$; its
\emph{location} does, through $b_H$. And blind cession does not close the interval: at $r^\ast$ and
$q=0$ the scenario band is $[\nbandloclsblind,\nbandhiclsblind]$ under class rating, against
$[\nbandlounins,\nbandhiunins]$ uninsured.

\subsection{Existence is not selection}

\begin{proposition}[Exact null-pooling condition]\label{prop:nullpool}
Fix a feasible pledge and the policy. There exists a perfect Bayesian equilibrium in which both types
choose the null contract if and only if
\begin{equation}
\mu_H\,G\ \ge\ \min\{G,\ t(D-c_0)\Delta\rho\}-\ke-b_HD.
\label{eq:nullpool}
\end{equation}
This condition can hold simultaneously with both strict inequalities in \eqref{eq:icir}.
\end{proposition}

The stylised example in Appendix~A (with $\rhoF=.2$, $\rhoH=.8$, $x_b=.5$, $v=5/6$, $D=1$) has
$G=\nexG$, $\ke=\nexkeff$, $b_H=\nexbH$, $b_F=\nexbF$: the competent type strictly prefers the
certificate (payoff \nexHsep) and the mimic strictly prefers the null ($\nexFmim$), and yet null
pooling is supported by an off-path belief $F$ under which both types prefer the pooled null receipt
\nexnullbid\ to their certificate deviations (\nexHdev\ and \nexFdev). Every availability statement
in this paper is therefore existential---a menu \emph{can} support the separating equilibrium---and
the ticket bands and masses describe available menus, not uniquely predicted take-up.

\subsection{Who keeps the certification rent}

\begin{figure}[t]
\centering\includegraphics[width=\textwidth]{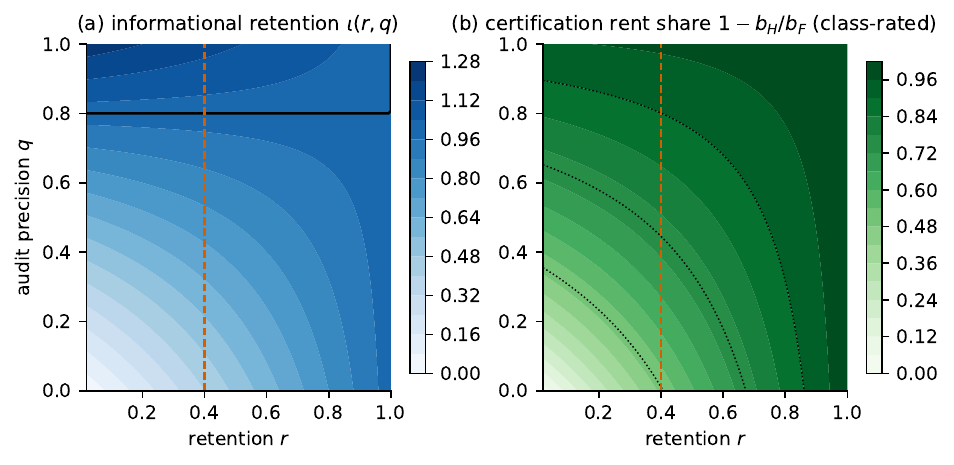}
\caption{\textbf{Informational retention and the certification rent.} (a) $\iota(r,q)$; the black
contour is $\iota=1$, above which loaded audited cover is more type-sensitive than self-retention.
The dashed line is $r^\ast=W/(W+K)$. (b) The share $\eta$ of the certification rent kept by the
competent provider under class rating; dotted contours at 50, 75 and 90\%.}
\label{fig:retention}
\end{figure}

On a separating menu the competent provider's gain from certifying at the smallest deterring pledge
is $(G-\ke)(1-b_H/b_F)$. The factor
\begin{equation}
\eta(r,q)\equiv1-\frac{b_H}{b_F}=\frac{\iota\Delta\rho}{\iota\Delta\rho+(1-r)\{\li x_H+a_I(1-q)(x_b-x_H)\}}
\label{eq:rent}
\end{equation}
is the share of the certification rent the competent provider keeps; the rest is the pooled load it
pays for the mimic's benefit. Under experience rating $b_H$ is the loading alone and $\eta$ is
\nrentexpblind\% even for a blind carrier. Under class rating $\eta$ rises from \nrentclsblind\% at
$q=0$ to \nrentclsfull\% at $q=1$: monitoring transfers the cross-subsidy back to the provider whose
competence the certificate is about. Figure~\ref{fig:retention}(b) maps $\eta$.

\section{Capacity: width and location}\label{sec:capacity}

Define the \emph{effective informational capacity} $\Chat(r,q)\equiv\iota(r,q)\Db(r)$. It is the
largest coefficient the policy can place on the difference between the two types' expected losses,
and every statement about who can be served is a statement about it and about the pooled load.

\begin{theorem}[Maximum incentive width]\label{thm:capacity}
For fixed $q\in[0,1]$ and $r^\ast=W/(W+K)$,
\begin{equation}
\max_{0\le r\le1}\Chat(r,q)=W+a_IqK,
\label{eq:capacity}
\end{equation}
attained at $r^\ast$, uniquely if $q>0$ and on $[r^\ast,1]$ if $q=0$. If $W+K\ge c_0$, the largest
width of the separating $G$-interval of Theorem~\ref{thm:binary} over feasible menus and retentions is
$t\Delta\rho\,(W+a_IqK)$; if $W+K<c_0$ no certificate is feasible.
\end{theorem}

\begin{figure}[t]
\centering\includegraphics[width=.62\textwidth]{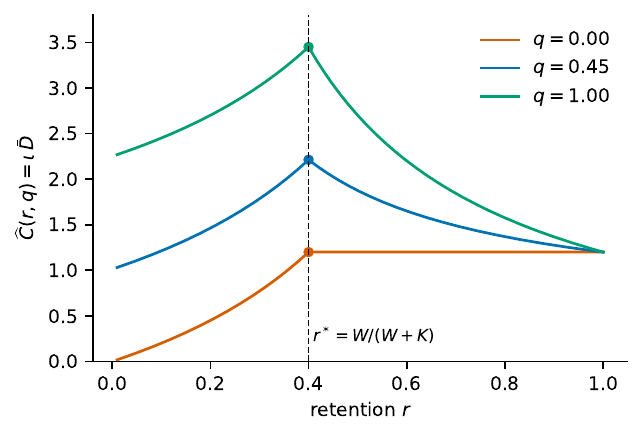}
\caption{\textbf{Effective informational capacity} $\Chat(r,q)=\iota\Db$ against retention. The
kink is at $r^\ast=W/(W+K)$ for every $q$; dots mark the closed form $W+a_IqK$. At $q=0$ the maximum
is a plateau at $W$ over $[r^\ast,1]$.}
\label{fig:chat}
\end{figure}

The mechanism at $q=0$ is an exact cancellation. Ceding a share $1-r$ multiplies the credible ceiling
by $1/r$ on the capital constraint; it multiplies the informational retention by $r$; and the width
of the separating interval scales as $\iota D$. A provider that cedes ninety percent of its exposure
to a blind carrier can post a pledge ten times larger, and the difference between what it pays and
what a mimic would pay is exactly one-tenth as sensitive to the pledge. Two readings follow. The
structure of the optimal programme---put exactly enough own capital behind the pledge that both
constraints bind---is invariant to how well the underwriter looks; monitoring changes the value of
the programme, not its design. And the marginal informational product of capacity is $a_Iq$: zero for
a blind carrier and linear thereafter. In the scenario $\Chat^\ast$ is \nChatblind\ blind,
\nChatbase\ at $q=\nPq$ and \nChatfull\ at $q=1$, against $W=\nPW$; Figure~\ref{fig:chat} shows the
kink at $r^\ast$ for every $q$.

\begin{theorem}[The largest certifiable ticket]\label{thm:frontier}
Suppose $u>0$ and $W+K\ge c_0$. Over all retentions and binary menus, the largest ticket for which a
separating menu exists is
\begin{equation}
v_{\rm top}(q)=\frac{\ke+t\{W\Delta\rho+K[a_I(qx_F+(1-q)x_b)-x_H]\}}{\zeta p_Au\Delta\rho},
\label{eq:frontier}
\end{equation}
attained at $r^\ast$ and $D=W+K$. At the fully deployed menu the participation and deterrence bounds
are
\begin{equation}
G^{\rm full}_{\rm lo}(q)=\ke+tK\{a_I[qx_H+(1-q)x_b]-x_H\},\qquad
G^{\rm full}_{\rm hi}(q)=G^{\rm full}_{\rm lo}(q)+t\Delta\rho\,(W+a_IqK).
\label{eq:full}
\end{equation}
\end{theorem}

\begin{corollary}[What blind capacity does]\label{cor:location}
At $q=0$, a unit of underwriter capacity raises both bounds in \eqref{eq:full} by
$t\{a_Ix_b-x_H\}=t\{\li x_H+a_I(x_b-x_H)\}$ and the width by nothing. Under experience rating
($x_b=x_H$) the shift is $t\li x_H$, the loading alone, and vanishes at actuarially fair prices.
Under class rating it carries the additional term $ta_I(x_b-x_H)>0$: the cross-subsidy the competent
provider pays for being pooled with the fringe, which makes the certificate costlier for both types
by the same amount and moves the served interval toward larger tickets.
\end{corollary}

\begin{figure}[t]
\centering\includegraphics[width=\textwidth]{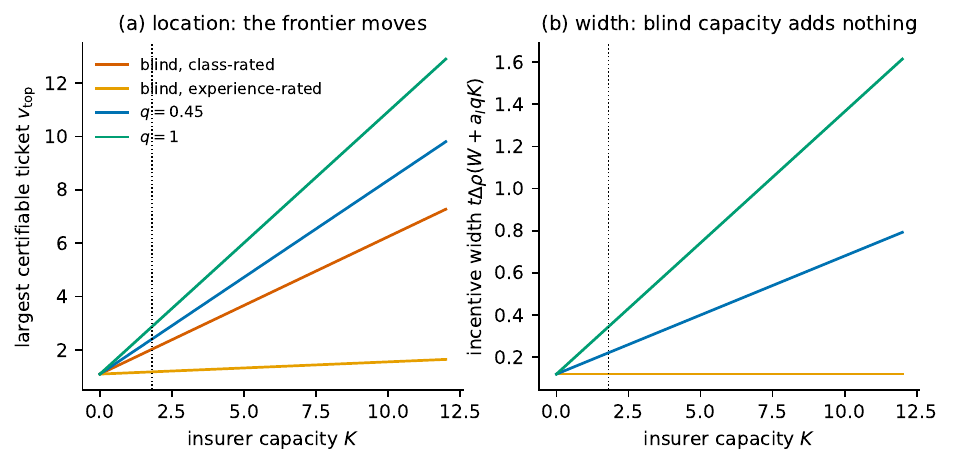}
\caption{\textbf{Location and width.} (a) The largest certifiable ticket $v_{\rm top}$ against
underwriter capacity $K$ at the capital-balancing retention, for a blind class-rated carrier, a blind
experience-rated carrier, and two audit precisions. (b) The width $t\Delta\rho(W+a_IqK)$ of the
separating interval: flat in $K$ for every blind carrier, linear with slope $t\Delta\rho a_Iq$
otherwise. The dotted line marks the scenario $K$.}
\label{fig:capacity}
\end{figure}

\begin{figure}[t]
\centering\includegraphics[width=\textwidth]{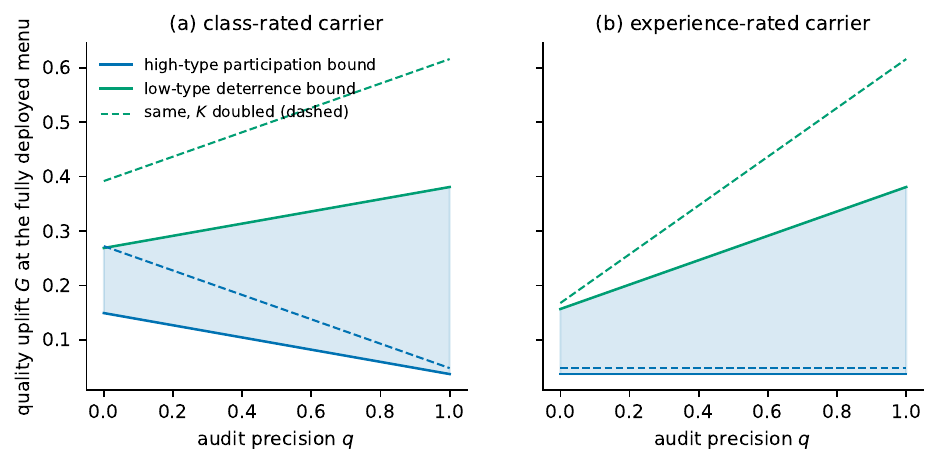}
\caption{\textbf{The incentive bands at the fully deployed menu} ($D=W+K$, $r=r^\ast$) against
audit precision. Monitoring lowers the competent provider's participation bound and raises the
mimic's deterrence bound. Doubling $K$ (dashed) shifts both bounds at $q=0$ by the same amount under
class rating (a) and by the loading alone under experience rating (b); the difference between them
widens only through $a_IqK$.}
\label{fig:bands}
\end{figure}

This is the corrected form of the capacity illusion, and it is sharper than the naive one. Unmonitored
capacity does not widen the range any one certificate discriminates. Whether it moves that range
depends on a rating convention that
insurance economists already track: an experience-rated blind carrier at fair prices leaves both the
width and the location alone, so third-party capacity is then informationally worthless in every
sense; a class-rated blind carrier relocates the market upward by exactly the cross-subsidy. In the
scenario the shift per unit of $K$ is \nshiftcls\ under class rating, of which \nshiftcrosssub\ is
the cross-subsidy and \nshiftloading\ the loading, against \nshiftexp\ under experience rating and
zero at fair loading. Raising $K$ from \nPK\ to 100 under a blind class-rated carrier moves the
largest certifiable ticket from \nvtopclsblind\ to \nvtopclsblindbigK, and the certificate that
serves that ticket discriminates the same $t\Delta\rho W$ of stakes around it that a certificate
did at $K=\nPK$; a full audit at $K=\nPK$ reaches \nvtopclsfull\ by widening.
Figure~\ref{fig:capacity} shows both objects against $K$, and Figure~\ref{fig:bands} the bounds
\eqref{eq:full} against $q$ under the two conventions.

\begin{remark}[Reach of one certificate versus the union of all certificates]\label{rem:union}
Two sets are easily conflated. The interval of Theorem~\ref{thm:binary} is what a \emph{given}
certificate $D$ separates; its width in $G$ is $t\iota\Delta\rho D$ and is bounded by
Theorem~\ref{thm:capacity}. The band \eqref{eq:tickets} of Corollary~\ref{cor:tickets} is the union
over all feasible $D$: its upper end uses $D=\Db$ and its lower end $D=c_0$, so it is
$(b_F\Db-b_Hc_0)/g$ wide and does grow in $K$ even at $q=0$---through $b_F\Db$, that is, through the
pooled load deterring the mimic at large pledges. The two statements are compatible. Under a blind
carrier each additional euro of capacity lets the institution place a certificate at higher stakes,
but every certificate it places, at any stakes, discriminates a range of width $t\Delta\rho W$;
covering the union requires one certificate per such range. Availability masses reported below are
unions and are labelled as such; the informational content of the cover is the per-certificate
width.
\end{remark}

\subsection{Comparative statics and the 2026 market}

\begin{corollary}[Exclusions and towers]\label{cor:exclusions}
At $r^\ast$, $\partial v_{\rm top}/\partial K=t\{a_I(qx_F+(1-q)x_b)-x_H\}/g$ and
$\partial(\text{width})/\partial K=t\Delta\rho a_Iq$. A generative-AI exclusion endorsement, a fall in
$K$, therefore narrows the range a certificate discriminates where underwriting inspects
human-in-the-loop practice and, where it does not, lowers the largest certifiable engagement without
narrowing that range---by the loading under experience rating and by the cross-subsidy under class
rating. Symmetrically, affirmative AI liability towers widen what a certificate discriminates only in
proportion to the diligence attached to them.
\end{corollary}

This is a testable heterogeneity. The natural reading of \citet{verisk2026} is that narrowing cover
shrinks the market for guaranteed professional services. The model says the shrinkage is concentrated
in the monitored segment, and that in the unmonitored class-rated segment the exclusion moves the
certification market toward smaller engagements---which, since the cross-subsidy is removed with the
capacity, is a gain for the provider on every ticket it still serves.

\begin{corollary}[Better machines narrow the band from below, and insurance cannot stop it]
\label{cor:betterai}
As the unassisted machine improves, $\rhoF$ rises and $\Delta\rho$ falls. Then
\[
v_{\rm lo}=\frac{\ke+b_Hc_0}{\zeta p_Au\,\Delta\rho}\ \longrightarrow\ \infty
\]
for every configuration $(r,q,K,x_b)$, because $\ke>0$ does not depend on the gap. No insurance arrangement offsets it: insurance acts on the provider's
cost of signalling and not on the gap being signalled.
\end{corollary}

\begin{remark}[The penalty doctrine: a third bound, abstracted from]\label{rem:doctrine}
\eqref{eq:cap} is the solvency pair. \citet{bauer2026threeceilings} shows there is a second bound
that no capital structure relaxes: the penalty doctrine caps the enforceable term at a
jurisdictional multiple $m$ of the loss, so the effective ceiling is $\min\{\Db(r),mv\}$. Under the
hypotheses of Corollary~\ref{cor:tickets} the feasibility test becomes $\max\{c_0,(G-\ke)/b_F\}\le
\min\{\Db,mv,(G-\ke)/b_H\}$. Where the doctrine binds it binds hardest on small tickets, since it
scales with $v$ and solvency does not, and it is untouched by insurance: Theorems~\ref{thm:capacity}
and~\ref{thm:frontier} describe what underwriting can do to the solvency leg, and the answer for the
doctrine leg is nothing.
\end{remark}

\section{Monitoring, disclosure, and the certification window}\label{sec:monitoring}

Everything so far treats the underwriter's information as private. It is not. Professional-indemnity
terms are disclosed in tenders, brokers publish ratings, and coverage schedules travel with
procurement packs. Let $d\in[0,1]$ be the share of what the underwriter learns that reaches clients,
and let the residual demand for the provider's own signal be $u=1-dq$, so that $g(q)=\zeta p_A(1-dq)
\Delta\rho$. This is an exogenous reduction in the value of the provider's own certificate---the
public shadow of the audit does part of its work---not a posterior in which the client learns the
audit's outcome; the population still contains both types with positive probability.

\subsection{The window}

Monitoring now has two opposed effects on the availability of a client-facing certificate. It lowers
$b_H$ and raises $b_F$, which moves both ends of the ticket interval \eqref{eq:tickets} favourably;
and it lowers $g$, which scales both ends upward and pushes the interval out of the mass of the ticket
distribution. Let $M(q;d)=F(v_{\rm hi}(q))-F(v_{\rm lo}(q))$ be the share of tickets for which some
separating menu exists at a fixed retention.

\begin{proposition}[Certification window]\label{prop:window}
Fix $r$ with $\Db(r)\ge c_0$. Then
\begin{equation}
M'(q)=f(v_{\rm hi})\,v_{\rm hi}'(q)-f(v_{\rm lo})\,v_{\rm lo}'(q),
\label{eq:Mprime}
\end{equation}
with $v_{\rm hi}'(q)>0$ for every $d\ge0$, and
$\operatorname{sign}v_{\rm lo}'(0)=\operatorname{sign}(d-d^\ast)$, where
\begin{equation}
d^\ast=\frac{c_0\,t\,a_I(1-r)(x_b-x_H)}{\ke+b_H(0)c_0}.
\label{eq:dstar}
\end{equation}
Hence (i) at $d=0$, $M$ is nondecreasing in $q$; (ii) for $d<d^\ast$, $M'(0)>0$; (iii) at $d=1$,
$M(q)\to0$ as $q\to1$, so that whenever $M'(0)>0$ an interior maximiser exists, and by continuity it
persists for $d$ in a neighbourhood of one. Under experience rating $d^\ast=0$ and the initial rise
comes from the deterrence bound alone.
\end{proposition}

\begin{figure}[t]
\centering\includegraphics[width=\textwidth]{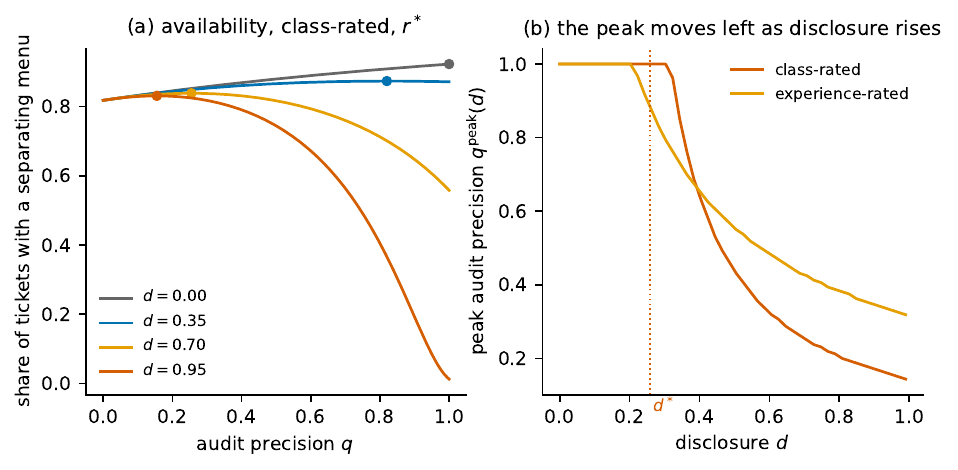}
\caption{\textbf{The certification window.} (a) Share of tickets with a separating menu against
audit precision, for four disclosure levels, class-rated, $r^\ast$; dots mark the maxima. (b) The
audit precision at which availability peaks, against disclosure, for both rating conventions; the
dotted line is $d^\ast$ from \eqref{eq:dstar}.}
\label{fig:window}
\end{figure}

The proposition states what can be proved: the sign of each boundary term, the threshold on
disclosure below which the market grows from both ends, and the existence of an interior peak at high
disclosure. The scenario supplies the shape. Under class rating the peak is at $q=1$ (the boundary)
for $d=0$, at \nqpeakclsdlow\ for $d=0.35$, \nqpeakclsdmid\ for $d=0.70$ and \nqpeakclsdhigh\ for
$d=0.95$, with $d^\ast=\ndstarcls$; under experience rating the corresponding peaks are
\nqpeakexpdlow, \nqpeakexpdmid\ and \nqpeakexpdhigh. The peak moves left as disclosure rises in every
one of \nrobn\ random parameter draws (online appendix). Figure~\ref{fig:window} shows the curves and
the peak locus. This is the informational-redundancy channel that \citet{bauer2026fallbacksignal}
found between competing providers, relocated to the vertical relationship, where it is sharper
because the underwriter's information is better than any rival's.

\subsection{Who pays for the audit}

The audit has a cost, $c_mq^2/2$, and somebody has to bear it. Consider the fully deployed menu at
$r^\ast$ with $D=\Db$. The underwriter's expected margin on the certified path is
\begin{equation}
m(q)=P_H-(1-r)tDx_H=tD(1-r)\{\li x_H+a_I(1-q)(x_b-x_H)\}=b_H(q)\,D,
\label{eq:margin}
\end{equation}
which is exactly the competent provider's net certificate cost: every euro of pooled load the
provider pays is a euro of margin the underwriter earns. Monitoring lowers $m$ under class rating
and leaves it unchanged under experience rating. Define the underwriter's private objective
$J_C(q)=m(q)-c_mq^2/2$, the joint provider--underwriter surplus over the ticket distribution
$J(q)=\int_{v_{\rm lo}}^{v_{\rm hi}}[G(v)-\ke]\,\dd F(v)-c_mq^2/2$, in which the transfer $b_HD$
cancels, and the client surplus over the served band $E(q)=\frac{1-\zeta}{\zeta}\int_{v_{\rm
lo}}^{v_{\rm hi}}G(v)\,\dd F(v)$.

\begin{proposition}[The monitoring market]\label{prop:monmarket}
Let $d=0$. (i) The underwriter acting alone chooses $q_P=0$ for every $c_m>0$. (ii) The joint
optimum satisfies $q_J>0$. (iii) $E$ is nondecreasing in $q$; if $J$ has a unique maximiser, every
maximiser of $J+E$ is at least $q_J$, and strictly larger if $q_J\in(0,1)$ and $E'(q_J)>0$. Hence
$0=q_P<q_J\le q_S$.
\end{proposition}

\begin{figure}[t]
\centering\includegraphics[width=.66\textwidth]{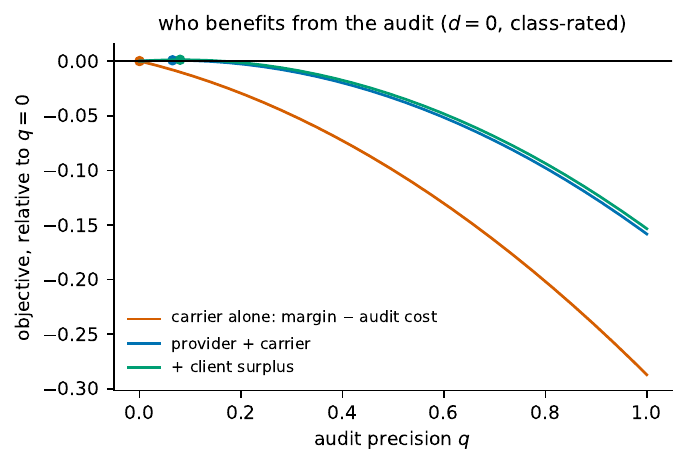}
\caption{\textbf{Who benefits from the audit.} The underwriter's own objective, the joint
provider--underwriter surplus, and the social objective including client surplus, each against audit
precision and relative to its value at $q=0$; class-rated, $d=0$, scenario values. Dots mark the
maximisers $q_P$, $q_J$, $q_S$.}
\label{fig:monitoring}
\end{figure}

The audit is worth nothing to the underwriter alone: under class rating it destroys the
cross-subsidy the underwriter was collecting, under experience rating it changes nothing on the
certified path, and it costs money either way. It is worth something to the provider, at the
intensive margin under class rating (the provider's willingness to pay per unit of precision,
$tDa_I(1-r)(x_b-x_H)=\nwtpcls$, equals the underwriter's margin loss) and at the extensive margin
under either convention (tickets that become certifiable). And it is worth more still to clients,
who capture $(1-\zeta)/\zeta$ of the quality uplift on every ticket the audit adds to the band. In
the scenario $q_J=\nqJ$ and $q_S=\nqS$; Figure~\ref{fig:monitoring} plots the three objectives. The
regulatory hook is specific: the intervention that widens the market is not more capacity and not a
higher liability cap, but a disclosure or documentation standard that makes underwriting diligence
contractible, so that the party who values it can buy it from the party who supplies it.

\subsection{Endogenous monitoring: global, not local}

A carrier that sees a provider retain a large share may rationally economise on audit; one that fears
adverse selection at high cession may do the opposite. Either way $q$ becomes a function of $r$ and
Theorem~\ref{thm:capacity}, which fixes $q$, is no longer immediate.

\begin{proposition}[Exact global test]\label{prop:global}
Let $q(\cdot)$ be continuous and $q_\ast=q(r^\ast)$. Then $r^\ast$ maximises $\Chat(r,q(r))$ if and
only if
\begin{equation}
a_IK[q(r)-q_\ast]\le W-\tfrac{Kr}{1-r}\ \ (0\le r<r^\ast),\qquad
Wq(r)\tfrac{1-r}{r}\le Kq_\ast\ \ (r^\ast<r\le1).
\label{eq:global}
\end{equation}
If $q$ is differentiable on each branch, $q'(r)\ge-1/[a_I(1-r)^2]$ on $(0,r^\ast)$ and
$q'(r)\le q(r)/[r(1-r)]$ on $(r^\ast,1)$ are sufficient, provided they hold over the whole branch.
\end{proposition}

Local slopes at the crossing are not enough: a smooth audit response that is flat at $r^\ast$ and
rises away from it can place the maximum elsewhere. The sufficient inequalities are wide---at the
scenario values monitoring may fall by more than two precision points per unit of retention before
the argument breaks---and they are an empirically checkable underwriting practice rather than a
technicality.

\section{Coverage and the stock it insures}\label{sec:coverage}

The type is produced by an engagement decision, so the insurance contract feeds back into how much
fallback capability exists at all. The classical expectation is moral hazard: cover dulls care. Here
that expectation is right for two specific reasons, and monitoring removes one of them.

\begin{proposition}[Two channels of erosion]\label{prop:invest}
Let a provider be able to become the competent type at cost $\kappa$, and let the menu, ticket and
policy be fixed with the separating equilibrium played. The return to investing is
$R(r,q)=G-\ke-b_H(r,q)D$, and
\begin{equation}
\frac{\partial R}{\partial r}=tD\,\big\{\li x_H+a_I(1-q)(x_b-x_H)\big\}\ \ge0 .
\label{eq:invest}
\end{equation}
Ceding exposure lowers the return to preserving the fallback through the loading, for every $q$ and
every rating convention, and through the cross-subsidy, under class rating only; the second channel
vanishes at $q=1$.
\end{proposition}

\begin{figure}[t]
\centering\includegraphics[width=\textwidth]{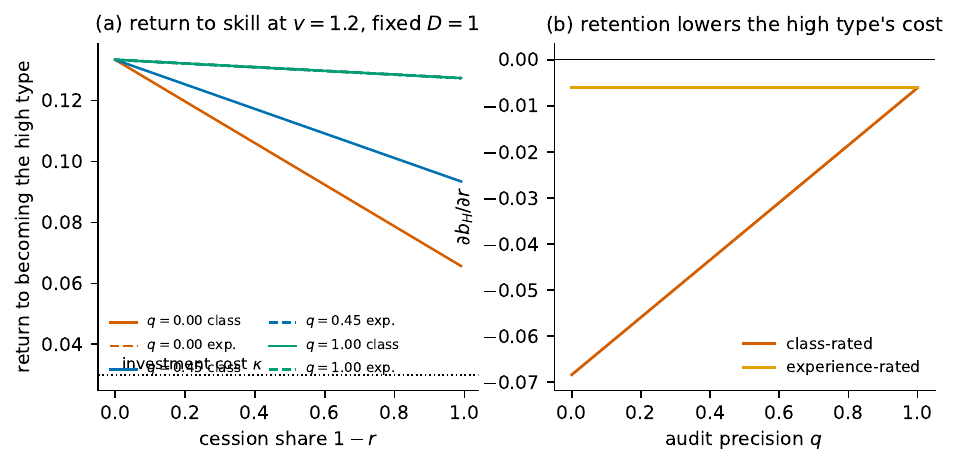}
\caption{\textbf{Cover and the incentive to preserve the fallback.} (a) The return to becoming the
competent type against the cession share, at a fixed certificate, for three audit precisions and both
rating conventions (solid: class-rated; dashed: experience-rated); the dotted line is the investment
cost $\kappa$. (b) The retention derivative of the competent provider's certificate cost against audit
precision: the gap between the two conventions is the cross-subsidy, closed by monitoring.}
\label{fig:investment}
\end{figure}

In the scenario the retention derivative of the return is \ndbHdrclsblind\ (in absolute value) under
a blind class-rated carrier and \ndbHdrloading\ under a full audit or under experience rating; the
cross-subsidy accounts for the difference. Figure~\ref{fig:investment} traces the return against the
cession share.

For a continuous skill stock with a differentiable rescue function, the same logic holds at the
margin. Holding $D$, $r$, $q$ fixed, the gross saving from raising skill is $-C_s(s,D)=tD\iota\rho'(s)$,
and its retention derivative is $tD\rho'(s)[1-a_Iq]$: at a point where $\rho'(s)>0$, retaining more
raises the marginal return to skill when $q<1/a_I$ and lowers it when $q>1/a_I=\nqswitch$. Beyond
that precision a loaded audited premium is more type-sensitive than self-retention, which is the
signalling-subsidy property of Lemma~\ref{lem:decomp} seen from the incentive side. We do not derive
an abandonment threshold or a bang-bang engagement policy from this: doing so requires a global
comparison of the intertemporal objective at every feasible engagement level, which the sign of an
endpoint derivative does not deliver. The learning--erosion transition of \citet{singh2026fallback}
enters as an accounting illustration of how the stock moves, projected onto its domain, not as a
solved dynamic programme.

\section{Coarse certification with many types}\label{sec:coarse}

Nothing above depends on there being two types; what matters is that there is one certificate. Let
skill $s$ be distributed on $[s_F,s_H]$ with an atomless distribution $F_s$, let $\rho(s)$ be
continuous and strictly increasing, and keep the binary menu, the policy and a class-rated reference
$x_b=1-E[\rho]$. A \emph{cutoff profile} $\hat s$ has types $s\ge\hat s$ certify and types $s<\hat
s$ choose the null contract; Bayes' rule gives $\rho_C(\hat s)=E[\rho(s)\mid s\ge\hat s]$ after the
certificate and $\rho_N(\hat s)=E[\rho(s)\mid s<\hat s]$ after the null.

\begin{proposition}[Cutoff equilibria]\label{prop:coarse}
Suppose $\iota>0$. (i) Given any beliefs, the certificate-minus-null payoff is strictly increasing in
$s$, so every best response is a cutoff. (ii) Define
\begin{equation}
h(\hat s)=\chi_v[\rho_C(\hat s)-\rho_N(\hat s)]+t(D-c_0)[1-\rho_C(\hat s)]-tD[\iota(1-\rho(\hat
s))+\ap]-\kd .
\label{eq:cutoff}
\end{equation}
An interior cutoff $\hat s\in(s_F,s_H)$ is a perfect Bayesian equilibrium if and only if
$h(\hat s)=0$; $h$ is continuous, and if $\lim_{\hat s\downarrow s_F}h<0<\lim_{\hat s\uparrow s_H}h$
an interior cutoff equilibrium exists. (iii) The two-type model is the special case of a two-point
distribution, with $\Delta\rho$ read as $\rho_C-\rho_N$.
\end{proposition}

\begin{figure}[t]
\centering\includegraphics[width=.66\textwidth]{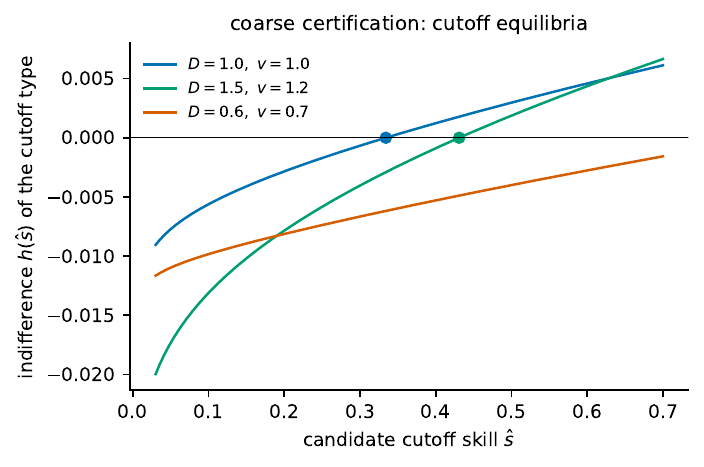}
\caption{\textbf{Cutoff equilibria with a continuum of types.} The indifference function
\eqref{eq:cutoff} of the marginal type for three (ticket, pledge) pairs; skill uniform on
$[s_F,s_H]$, class-rated reference, scenario values. A root is a cutoff equilibrium; a function
that stays negative supports only the all-null outcome.}
\label{fig:coarse}
\end{figure}

At the scenario values with $D=1$ and $v=1$ the cutoff is $\hat s=\ncutoffDoneVone$, a rescue
probability of \nrhocutoffDoneVone; at $D=1.5$, $v=1.2$ it is \ncutoffDbigVbig. Multiple roots are
possible and the proposition claims existence, not uniqueness. Figure~\ref{fig:coarse} plots
$h$. The point of the extension is that the certificate is a \emph{coarse} instrument by nature: a
continuous pledge schedule that separates every type does not survive the null option and the
client's expected recovery (Appendix~B), and a coarse certificate is also what keeps the underwriter's
inference coarse, so that its audit has something to add.

\section{Ex-ante screening: the deep-pockets objection}\label{sec:objections}

Drop Assumption~\ref{as:noscreen}. Real carriers decline submissions: even one that never audits may
screen at the point of sale, so that being granted a large limit is itself informative. Suppose the
underwriter screens with precision $q_0$ and that the grant is observable to clients.

\begin{proposition}[Screening is disclosure, not retention]\label{prop:screen}
Ex-ante screening does not enter $\iota$, $\lp$ or the cost coefficients \eqref{eq:bH}--\eqref{eq:bF}:
it affects who is offered cover, not how the ceded leg is priced against the realised type. It enters
the residual demand for the provider's own signal, $u=(1-dq)(1-q_0)$, and therefore scales both ends
of the ticket interval \eqref{eq:tickets} by $1/(1-q_0)$ without changing its width in $G$-space.
Theorems~\ref{thm:capacity} and~\ref{thm:frontier} are unaffected as statements about separation
through the pledge.
\end{proposition}

The objection says that a screened limit is informative; the model agrees, and places that
information in the channel where the paper already shows it does damage. A market in which carriers
screen hard and limit grants are public is a market in which the underwriter has become the certifier
and the provider's own pledge is redundant---the high-disclosure corner of Figure~\ref{fig:window},
reached from a different direction. What we have not shown is that a screened limit \emph{separates}:
that would require modelling the carrier's acceptance rule as an equilibrium object, with the
provider choosing whether to submit. Nor have we found a model in which limit size certifies quality
to the insured's own customers through ex-ante screening; the closest results certify to investors
through the firm's own purchase decision \citep{thakor1982exploration,gracerebello1993financing}.
A screening-and-certification model of underwriting capacity is a paper, not a footnote.

\section{What a verification record can and cannot establish}\label{sec:records}

The analytical results assume clients hold correct equilibrium beliefs and the audit is a scalar
precision applied to a known type. In practice clients see a public record of verified rescues and
verified failures, censored by verification, and the underwriter audits a sample. We separate two
questions.

\begin{proposition}[Censored records]\label{prop:statistics}
Conditional on an AI-down event, let a rescue be recorded with probability one and an unrescued
failure with probability $\varphi(1-\rho)$. Among recorded events the success fraction converges to
$z=\rho/[\rho+\varphi(1-\rho)]\ge\rho$, with equality only at $\varphi=1$ or at the boundary. Let
$N$ recorded events be i.i.d.\ with success probability $z$, $\widehat z_N$ their sample proportion,
and $\widehat\rho_N=\varphi\widehat z_N/[1-(1-\varphi)\widehat z_N]$. Then
$\widehat\rho_N-\rho=O_p(N^{-1/2})$, and the same rate holds for any cost coefficient affine in
$\rho$.
\end{proposition}

At the scenario values the record overstates the competent type's rescue probability as
$z=\nzobsH$ against $\rhoH=\nrhoH$; the inverse map recovers it exactly. This is a statistical
statement about estimating costs from records, and we state what it is not: if clients use the record
to change their bids, provider incentives must be recomputed from those bids, and a report that
blends a record with a claimed type is not truth-telling---the derivative of the provider's payoff at
the truth carries a term $-w\chi\rho'(s)<0$ for any positive record weight $w$, so a provider gains by
under-reporting. A corrected record does not restore that equilibrium; Theorem~\ref{thm:binary} does
not need it.

\begin{figure}[t]
\centering\includegraphics[width=\textwidth]{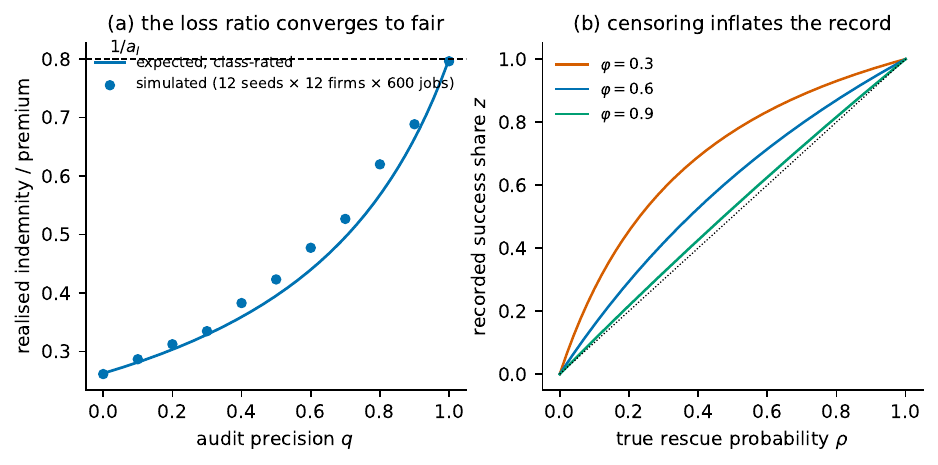}
\caption{\textbf{Records and the loss ratio.} (a) The underwriter's realised loss ratio on the
certified path against audit precision, expected and simulated (12 seeds $\times$ 12 firms $\times$
600 jobs); the dashed line is $1/a_I$. (b) The recorded success share against the true rescue
probability for three verification probabilities: censoring inflates the record.}
\label{fig:market}
\end{figure}

What a stochastic market does test is the pricing rule. Under class rating the underwriter's
realised indemnity over premium on the certified path is $x_H/[a_I(qx_H+(1-q)x_b)]$: \nLRanblind\
for a blind carrier, rising to $1/a_I=\nLRfair$ under a full audit. A simulation with realised
failures, censored verification and audits drawn at random reproduces it to sampling error
(Figure~\ref{fig:market}). The blind carrier's low loss ratio is not a profit in the usual sense: it
is the cross-subsidy of Corollary~\ref{cor:location}, collected from the competent provider, and it
is the same object that Proposition~\ref{prop:monmarket} says the carrier will not give up on its
own.

\section{Taking it to data}\label{sec:data}

The model's distinctive predictions are about underwriting practice, which is observable, and they
separate cleanly from a pure willingness-to-pay story. We sketch two designs; neither is run here,
and Table~\ref{tab:falsify} lists each claim with what would falsify it.

\paragraph{Design 1: archival, exploiting the 2026 exclusions.} The generative-AI exclusion
endorsements attaching to CGL renewals from January 2026 \citep{verisk2026} are a shock to $K$ with
plausibly exogenous timing relative to any individual professional firm.
Corollary~\ref{cor:exclusions} predicts a heterogeneity: among firms whose carriers conduct
human-in-the-loop diligence, the spread of engagement sizes served at a given guarantee level should
narrow; among firms whose carriers underwrite on revenue and claims history, the largest guaranteed
engagement should fall while that spread does not narrow, and the fall should be larger where cover
is class-rated. A uniform contraction across the two groups is what a pure capacity story predicts and
what would falsify the model. The required data are the retained exposure, the collectible limit,
the rating basis, and the distribution of guaranteed engagement sizes before and after the
endorsement; coverage limits alone are not sufficient.

\paragraph{Design 2: choice-based conjoint on the procurement side.} A preregistered CBC fielded on
B2B procurement decision-makers, with attributes: liability commitment (none / $1\times$ / $5\times$ /
$20\times$ fee); who bears it (self-retained / insured, carrier named / insured, carrier
undisclosed); underwriting diligence (none stated / annual questionnaire / audited exception-handling
drill); human-fallback guarantee (none / named reviewer / qualified reviewer with documented record);
price. The model's sharpest prediction is an interaction: willingness to pay for the liability
attribute should increase in stated underwriting diligence, and the fallback-guarantee partworth
should \emph{decrease} in disclosed diligence, which is the redundancy channel of
Proposition~\ref{prop:window}. Both partworths positive and additive falsifies the window. Powering
should be done against \citet{mccoll2019guarantees} for the guarantee attribute.

\begin{table}[t]
\centering\small
\caption{Claims and what would falsify them.}
\label{tab:falsify}
\begin{tabularx}{\textwidth}{@{}lcX@{}}
\toprule
Claim & Where & What would falsify it \\
\midrule
Cost difference requires $\iota>0$ (P\ref{prop:sc}) & \S\ref{sec:eqm} &
Insurer loss data showing indemnity uncorrelated with retention and diligence, conditional on exposure \\
Width is $t\Delta\rho(W+a_IqK)$ (T\ref{thm:capacity}) & \S\ref{sec:capacity} &
Firms raising ceded limits with unchanged diligence and serving a \emph{wider} range of guaranteed engagements \\
Blind capital relocates (C\ref{cor:location}) & \S\ref{sec:capacity} &
No difference between class- and experience-rated firms in the size of guaranteed engagements after a capacity shock \\
Cross-subsidy to the rent (eq.~\ref{eq:rent}) & \S\ref{sec:eqm} &
Certified providers' net margin on guarantees unrelated to carrier diligence under class rating \\
Two channels of erosion (P\ref{prop:invest}) & \S\ref{sec:coverage} &
Deskilling correlated with coverage ratio irrespective of carrier diligence and rating basis \\
Certification window (P\ref{prop:window}) & \S\ref{sec:monitoring} &
Guarantee provision monotone in disclosed diligence \\
Under-monitoring (P\ref{prop:monmarket}) & \S\ref{sec:monitoring} &
Carriers supplying audits unpaid by insureds under class rating \\
\bottomrule
\end{tabularx}
\end{table}

\section{Discussion}\label{sec:disc}

\subsection{What the model says to whom}

\paragraph{Providers.} Pledge capacity is a competitive asset, but only the audited part of it widens
the range of engagements a certificate can discriminate. The design of the programme is invariant: retain the share $r^\ast=W/(W+K)$
that makes both constraints bind. What is not invariant is which carrier writes it and on what basis.
Between two towers of identical limit and price, the one attached to an audit of exception-handling
practice is worth strictly more; and between two blind towers, the experience-rated one leaves the
provider its certification rent while the class-rated one takes half of it (\nrentclsblind\% kept in
the scenario). A provider that buys limits from a class-rated carrier that does not audit is paying
a cross-subsidy that carries no information, losing the small engagements it could have certified,
and weakening its own reason to keep people in the loop.

\paragraph{Underwriters.} The audit is under-supplied relative to the surplus it creates
(Proposition~\ref{prop:monmarket}), because the carrier captures the cross-subsidy it would destroy
and not the certification value it would create. A carrier that could publish its diligence standard
and price it---sell the certification value rather than give it away---would internalise more of it.
That points at a product that does not currently exist: a professional-indemnity policy whose
diligence tier is a disclosed, contractible attribute of the insured, priced separately from
capacity.

\paragraph{Regulators.} Three levers appear in the comparative statics and they are not the ones
under discussion. Documentation standards that make failures attributable raise $\varphi$; procedural
cost reduction lowers $c_0$; and a disclosure standard for underwriting diligence raises the return
to $q$ and makes it contractible. Raising liability caps or subsidising capacity does nothing for the
market's ability to discriminate competence where carriers do not look
(Theorem~\ref{thm:capacity}), and under class rating it moves that market away from the small
engagements where preserved competence matters most per euro. Since Regulation (EU) 2026/1744 defers
rather than creates a civil-liability channel \citep{eu2026omnibus}, and the AI Liability Directive
was withdrawn \citep{eu2025aild}, the market's ability to price preserved competence rests on private
contract and on the insurance that stands behind it---which is exactly where the model locates the
binding constraint.

\subsection{Limitations}

\emph{The certificate is binary and the menu is institutional.} Every equilibrium claim is
conditional on a two-contract menu with the policy fixed before the type is drawn. Providers cannot
post further terms, choose retentions type by type, or set private prices. Section~\ref{sec:coarse}
shows that the type space can be a continuum; the menu cannot, and Appendix~B says why. An
unrestricted-menu equilibrium is an open problem, and we do not claim that our conclusions survive
it unchanged.

\emph{Existence is not selection.} Proposition~\ref{prop:nullpool} shows that strict separating
incentives coexist with null pooling. The availability bands and masses describe menus that
\emph{can} support certification, not predicted take-up. Selection requires a refinement or an
experiment.

\emph{The rescue technology is imposed.} $\rho(s)$ is a functional-form assumption with no direct
empirical calibration. That a human review layer catches only a fraction of errors is documented
\citep{maksymov2024audit}; that readiness decays with disuse is supported from the automation-safety
side \citep{dixit2016takeover,bainbridge1983ironies}; what no study provides is a catch rate as a
function of maintained practice.

\emph{Monitoring is a scalar and the audit is unbiased.} Real diligence is a bundle with its own
incentive problems. A systematic bias in the audit posterior enters $\ap$, not $\iota$: it is a
second pooled load, and it tightens every result in the direction of blindness.

\emph{The tariff references are anchors, not a theory of rating.} We take the rating convention as
given by the underwriter's book. A carrier with market power would price-discriminate on type; a
competitive class-rated book with endogenous composition would drift toward experience rating as
non-certifying providers leave it. Both are extensions we name rather than take.

\emph{Verifiability is a scalar.} \citet{bauer2026threeceilings} shows that provability declines as
model error becomes correlated across systems sharing foundation models
\citep{goel2025greatmodels,kim2025correlated,kleinberg2021monoculture}. Every boundary here scales in
$\varphi$, so that mechanism transmits one-for-one; one thing insurance adds is that the
underwriter's audit is itself a verification technology, and if it runs on the same models as the
insured's production---a question no current policy form asks---then $q$ inherits the same blind
spots.

\emph{No dynamic underwriting.} The natural formulation of the missing dynamics is a continuous-time
game in which the underwriter adjusts capacity and audit intensity against an observable diffusion of
the provider's capability, with withdrawal of cover as a stopping boundary; the machinery exists
\citep{sannikov2008continuous,demarzo2006optimal} and has not, as far as we can find, been applied to
joint control of coverage and monitoring against a quality diffusion.

\subsection{Conclusion}

Once the artifact stops carrying information, the market for expert services becomes a market for
the capability to catch the machine, and that capability can only be certified by a commitment whose
cost falls as the capability rises. Such a commitment needs a balance sheet behind it, and the balance
sheet belongs to someone who can look. Whether that party looks determines whether the commitment
separates at all, how wide the market it serves can be, and whether the commitment erodes the very
capability it certifies. Whether that party prices on the competent provider or on the class
determines whether unmonitored capital leaves the market where it is or moves it away from the
engagements that need it most. And because what the underwriter learns leaks, the commitment is
worth making only in a window between a carrier that does not look and a carrier that tells everyone
what it saw.

The two developments reshaping this market pull the wrong way. Capacity is growing, which does
nothing for discrimination without diligence and relocates the market under class rating; and
diligence, where it exists, is disclosed, which makes the provider's own signal redundant. The party
best placed to widen the window is the one that captures the smallest share of the gain---and, under
class rating, the one that loses money by widening it.

\newpage
\appendix
\section{Proofs}\label{app:proofs}

Throughout, Assumptions~\ref{as:reg}--\ref{as:commit} hold and $\ap=\lp x_b$.

\paragraph{Proof of Lemma~\ref{lem:decomp}.}
Expected retained damages are $rtDx_T$. Adding \eqref{eq:premium},
$C_T=tD\{rx_T+a_I(1-r)[qx_T+(1-q)x_b]\}=tD\{[r+a_I(1-r)q]x_T+a_I(1-r)(1-q)x_b\}$, which is
\eqref{eq:cost}. \hfill$\square$

\paragraph{Proof of Proposition~\ref{prop:sc}.}
From \eqref{eq:cost}, $C_F-C_H=tD\iota(x_F-x_H)=tD\iota\Delta\rho$, strictly increasing in $D$ iff
$\iota>0$. At $\iota=0$ the two costs coincide, so \eqref{eq:icir} can hold only with $b_F=b_H$, i.e.
at equality. \hfill$\square$

\paragraph{Proof of Theorem~\ref{thm:binary}.}
Under the proposed strategies both messages occur with positive probability, so beliefs $F$ after the
null and $H$ after the certificate follow from Bayes' rule and bids are \eqref{eq:bid} at those
beliefs. The competent type's certificate revenue is $G+t(D-c_0)x_H$ and its cost $C_H+\kd$; using
\eqref{eq:cost}, $G+t(D-c_0)x_H-tD(\iota x_H+\ap)-\kd=G-\ke-b_HD$ with $b_H$, $\ke$ as in
\eqref{eq:bH}. Its only alternative is the null contract, worth zero at belief $F$; optimality is the
left inequality of \eqref{eq:icir}. A mimic receives the same revenue and pays $C_F+\kd$, giving
$G-\ke-b_FD$ with $b_F=b_H+t\iota\Delta\rho$; its only alternative is the null at belief $F$, worth
zero, so optimality is the right inequality. The action set contains exactly these two choices, so
the conditions are necessary and sufficient; buyers bid their values, the client pursues since
$D\ge c_0$, and the underwriter's path margin \eqref{eq:margin} is nonnegative. Strictness follows
from strict inequalities. \hfill$\square$

\paragraph{Proof of Corollary~\ref{cor:tickets}.}
Rearranging \eqref{eq:icir} in $D$ gives the pledge interval. Necessity of \eqref{eq:tickets}: the
left inequality with $D\ge c_0$ gives $G\ge\ke+b_Hc_0$; the right with $D\le\Db$ gives $G\le\ke+
b_F\Db$. Sufficiency: choose $D_\ell=\max\{c_0,(G-\ke)/b_F\}$; the upper ticket bound gives
$D_\ell\le\Db$; if $D_\ell=c_0$ the lower ticket bound gives participation, otherwise $G-\ke=b_FD_\ell
\ge b_HD_\ell$; deterrence holds by construction. \hfill$\square$

\paragraph{Proof of Proposition~\ref{prop:nullpool}.}
On the proposed path both types receive the null bid at the prior, $\mu_HG$, and incur no certificate
cost. The unobserved certificate may carry any posterior mean $\widehat\rho\in[\rhoF,\rhoH]$; its bid
\eqref{eq:bid} is affine in $\widehat\rho$, so the minimum over admissible beliefs is attained at an
endpoint. Since $C_F-C_H=tD\iota\Delta\rho\ge0$, the competent type has the larger deviation payoff at
every common bid. At belief $H$ that payoff is $G-\ke-b_HD$; at belief $F$ the bid is
$t(D-c_0)x_F$, and $t(D-c_0)x_F-C_H-\kd=t(D-c_0)\Delta\rho+[t(D-c_0)x_H-C_H-\kd]=t(D-c_0)\Delta\rho
-\ke-b_HD$. Hence the smallest attainable value of the larger deviation payoff is the right side of
\eqref{eq:nullpool}. If the inequality holds, assign the minimising endpoint belief; neither type
gains by deviating, Bayes' rule holds on the null path, and the strategies and beliefs form a PBE.
Conversely, if it fails, the competent type gains at every admissible off-path belief. \hfill$\square$

\paragraph{The coexistence example.} Take $\rhoF=.2$, $\rhoH=.8$, $x_b=.5$, $p_A=.25$, $\varphi=.6$,
$\zeta=.8$, $v=5/6$, $u=1$, $r=.4$, $q=.45$, $\li=.25$, $W=1.2$, $K=1.8$, $D=1$, $c_0=.25$, $\kd=.02$,
$\mu_H=.5$. Then $t=.15$, $\iota=\nexiota$, $G=\nexG$, $\ke=\nexkeff$, $b_H=\nexbH$, $b_F=\nexbF$.
Separation: $\nexIClo\le G\le\nexIChi$, both strict; payoffs $\nexHsep$ for $H$ and $\nexFmim$ for a
mimic. Null pooling: the null bid at the prior is $\nexnullbid$; assign the off-path certificate
belief $F$, bid $\nexbidF$; the deviation payoffs are $\nexHdev$ ($H$) and $\nexFdev$ ($F$), both
below $\nexnullbid$; condition \eqref{eq:nullpool} reads $\nexnullbid\ge\nexrhs$. Exact rational
arithmetic reproduces every value.

\paragraph{Proof of Theorem~\ref{thm:capacity}.}
The two capital bounds cross at $r^\ast$. For $0\le r\le r^\ast$, $\Db=K/(1-r)$ and
$\Chat=K\{r/(1-r)+a_Iq\}$, with derivative $K/(1-r)^2>0$. For $r^\ast\le r\le1$, $\Db=W/r$ and
$\Chat=W\{1+a_Iq(1-r)/r\}$, with derivative $-Wa_Iq/r^2$, strictly negative for $q>0$ and zero for
$q=0$. At the crossing the value is $W+a_IqK$. The $G$-interval in \eqref{eq:icir} has width
$(b_F-b_H)D=t\iota\Delta\rho D$, largest at $D=\Db$; adding $rD\le W$ and $(1-r)D\le K$ gives
$D\le W+K$, so the crossing is feasible iff $W+K\ge c_0$. \hfill$\square$

\paragraph{Proof of Theorem~\ref{thm:frontier}.}
The upper bound in \eqref{eq:icir} at $D=\Db$ is $\ke+b_F\Db$. Put $Z=a_I(qx_F+(1-q)x_b)$ so that
$b_F/t=rx_F+(1-r)Z-x_H$. On the insurer-constrained branch $b_F\Db/t=K\{\Delta\rho/(1-r)-(x_F-Z)\}$,
increasing in $r$; on the own-capital branch $b_F\Db/t=W\{x_F-Z+(Z-x_H)/r\}$, with derivative
$-W(Z-x_H)/r^2\le0$ because $x_F\ge x_b\ge x_H$ and $a_I\ge1$. Thus $r^\ast$ attains
$t\{W\Delta\rho+K(Z-x_H)\}$; the menu is feasible and satisfies participation since $b_F\ge b_H$;
dividing by $g>0$ gives \eqref{eq:frontier}. The bounds \eqref{eq:full} follow from
\eqref{eq:bH}--\eqref{eq:bF} at $r^\ast$, $D=W+K$, using $(1-r^\ast)(W+K)=K$. \hfill$\square$

\paragraph{Proofs of Corollaries~\ref{cor:location}--\ref{cor:betterai}.}
Immediate from \eqref{eq:full}, \eqref{eq:frontier} and \eqref{eq:tickets}: at $q=0$,
$\partial G^{\rm full}_{\rm lo}/\partial K=\partial G^{\rm full}_{\rm hi}/\partial K=t(a_Ix_b-x_H)$;
the width derivative is $t\Delta\rho a_Iq$; and $v_{\rm lo}\ge\ke/(\zeta p_Au\Delta\rho)\to\infty$
as $\Delta\rho\to0$. \hfill$\square$

\paragraph{Proof of Proposition~\ref{prop:window}.}
With $u=1-dq$, $v_{\rm hi}=(\ke+b_F\Db)/g_0(1-dq)$ and $v_{\rm lo}=(\ke+b_Hc_0)/g_0(1-dq)$, where
$g_0=\zeta p_A\Delta\rho$. Differentiating, $\partial_qb_F=ta_I(1-r)(x_F-x_b)\ge0$ and
$\partial_qb_H=-ta_I(1-r)(x_b-x_H)\le0$. Hence $v_{\rm hi}'=[\Db\partial_qb_F(1-dq)+d(\ke+b_F\Db)]/
g_0(1-dq)^2>0$ for $d\ge0$. Similarly $\operatorname{sign}v_{\rm lo}'(q)=\operatorname{sign}
[c_0\partial_qb_H(1-dq)+d(\ke+b_Hc_0)]$; at $q=0$ this is negative iff $d<c_0ta_I(1-r)(x_b-x_H)/
(\ke+b_H(0)c_0)=d^\ast$. Then \eqref{eq:Mprime} is the chain rule; (i) at $d=0$, $v_{\rm lo}'\le0
\le v_{\rm hi}'$ so $M'\ge0$; (ii) at $d<d^\ast$ both terms of \eqref{eq:Mprime} are positive at
$q=0$; (iii) at $d=1$ both endpoints diverge as $q\to1$, so $M\to0<M(0)$, and with $M'(0)>0$ a
maximiser on $[0,1]$ is interior; $M$ is continuous in $(q,d)$, so an interior maximiser persists in
a neighbourhood of $d=1$. Under experience rating $x_b=x_H$ gives $\partial_qb_H=0$ and $d^\ast=0$.
\hfill$\square$

\paragraph{Proof of Proposition~\ref{prop:monmarket}.}
(i) $m(q)=b_H(q)D$ is nonincreasing in $q$ by $\partial_qb_H\le0$, and $c_mq^2/2$ is increasing, so
$J_C$ is maximised at $0$. (ii) At $d=0$, $J'(q)=[G(v_{\rm hi})-\ke]f(v_{\rm hi})v_{\rm hi}'-
[G(v_{\rm lo})-\ke]f(v_{\rm lo})v_{\rm lo}'-c_mq$. Now $G(v_{\rm hi})-\ke=b_F\Db>0$, $G(v_{\rm lo})
-\ke=b_Hc_0\ge0$, $v_{\rm hi}'>0$ and $v_{\rm lo}'\le0$, so $J'(0)>0$ and the maximiser is positive.
(iii) At $d=0$ the band expands in $q$ and the integrand $G\ge0$, so $E$ is nondecreasing. For
$q<q_J$, uniqueness gives $J(q)<J(q_J)$ and monotonicity $E(q)\le E(q_J)$, so $q$ does not maximise
$J+E$; existence follows from compactness and continuity. If $q_J\in(0,1)$ is interior, $J'(q_J)=0$
and $(J+E)'(q_J)=E'(q_J)>0$, so $q_J$ is not a social maximiser. \hfill$\square$

\paragraph{Proof of Proposition~\ref{prop:global}.}
Substitute each branch of $\Chat$ from the proof of Theorem~\ref{thm:capacity} into
$\Chat(r,q(r))\le W+a_IKq_\ast$; rearranging gives \eqref{eq:global}, proving necessity and
sufficiency. Differentiating the branches gives $K\{1/(1-r)^2+a_Iq'(r)\}$ and
$Wa_I\{q'(r)(1-r)/r-q(r)/r^2\}$; the sufficient conditions make the first nonnegative and the second
nonpositive over the whole branch, and continuity extends the ordering to the endpoints.
\hfill$\square$

\paragraph{Proof of Proposition~\ref{prop:invest}.}
$R=G-\ke-b_HD$ with $b_H$ from \eqref{eq:bH}; $\partial b_H/\partial r=-t\{\li x_H+a_I(1-q)(x_b-
x_H)\}$, whence \eqref{eq:invest}. The bracket is $\li x_H$ when $x_b=x_H$ or $q=1$.
\hfill$\square$

\paragraph{Proof of Proposition~\ref{prop:coarse}.}
(i) Given beliefs, the certificate-minus-null payoff of type $s$ is $[\text{bid}_C-\text{bid}_N]-
tD[\iota(1-\rho(s))+\ap]-\kd$, strictly increasing in $s$ since $\rho$ is increasing and $\iota>0$;
a type prefers the certificate iff $s$ exceeds the indifferent type, so best responses are cutoffs.
(ii) At a cutoff profile $\hat s$ Bayes' rule gives $\rho_C(\hat s)$ and $\rho_N(\hat s)$, and the
marginal type $\hat s$ is indifferent iff $h(\hat s)=0$ with $h$ as in \eqref{eq:cutoff}; by (i)
all types above then strictly prefer the certificate and all below the null, so $h(\hat s)=0$ is
necessary and sufficient for an interior cutoff PBE. Conditional means of a continuous function
over an atomless distribution are continuous in the cutoff, so $h$ is continuous, and the
intermediate value theorem gives a root under the stated sign condition. (iii) A two-point
distribution at $\{s_F,s_H\}$ has $\rho_C=\rhoH$, $\rho_N=\rhoF$ at any interior cutoff, and
\eqref{eq:cutoff} evaluated at $\hat s=s_H$ and $\hat s=s_F$ reproduces the two inequalities of
\eqref{eq:icir}. \hfill$\square$

\paragraph{Proof of Proposition~\ref{prop:screen}.}
Screening affects the set of providers offered cover, not the premium \eqref{eq:premium} conditional
on cover, so $\iota$, $\lp$, $b_H$, $b_F$ are unchanged; the residual demand enters $g$
multiplicatively, scaling both ends of \eqref{eq:tickets}. \hfill$\square$

\paragraph{Proof of Proposition~\ref{prop:statistics}.}
Conditioning on being recorded gives $z$; $z\ge\rho$ since $\rho+\varphi(1-\rho)\le1$. The inverse
map is algebraic. The sample proportion is unbiased with variance $z(1-z)/N\le1/(4N)$, so
Chebyshev gives $\widehat z_N-z=O_p(N^{-1/2})$; the inverse map has derivative
$\varphi/[1-(1-\varphi)z]^2\le1/\varphi$ on $[0,1]$, and the mean-value inequality transfers the
rate; an affine transform multiplies the error by a constant. \hfill$\square$

\section{Why a continuous pledge schedule is not inherited}\label{app:continuum}

The companion \citet{bauer2026fallbacksignal} separates a
continuum of types by the schedule obtained from the local incentive-compatibility ODE integrated
from the bottom type at the pursuit floor,
\begin{equation}
D(s)=c_0+\frac{\chi}{t\iota}\log\frac{\iota(1-\rhoF)+\ap}{\iota(1-\rho(s))+\ap},
\label{eq:oldmenu}
\end{equation}
under a revenue $\chi\rho(\hat s)$ that does not include the client's expected recovery. Two things
are true of \eqref{eq:oldmenu} and one is not.

\emph{It has the within-menu report property.} For $\iota>0$ and $\rho'>0$ on a compact interval, the
report payoff $\chi\rho(\hat s)-tD(\hat s)[\iota(1-\rho(s))+\ap]-\kd$ has derivative
$\chi\rho'(\hat s)\iota[\rho(s)-\rho(\hat s)]/[\iota(1-\rho(\hat s))+\ap]$, positive below the true
type and negative above it: truth is the best report among the admitted reports.

\emph{It is not a participation-proof equilibrium.} If the null pledge carries belief $s_F$, the
lowest type loses $tc_0[\iota(1-\rhoF)+\ap]+\kd>0$ by posting $c_0$ rather than nothing---
$\nappBloss$ at the scenario values with $r=r^\ast=.4$, $q=.45$, class rating---and by continuity so do types near it. A
within-menu maximum cannot repair a profitable deviation to a contract outside the menu, and clipping
the schedule at a cap does not specify the beliefs of the resulting pool.

\emph{It does not survive the client's expected recovery.} Add $t[D(\hat s)-c_0][1-\rho(\hat s)]$ to
revenue, as \citet{bauer2026lastsignal} and Section~\ref{sec:model} require. The truthful first-order
condition becomes
\begin{equation}
t\{(\iota-1)(1-\rho(s))+\ap\}D'(s)=[\chi-t(D(s)-c_0)]\rho'(s).
\label{eq:newfoc}
\end{equation}
In the uninsured case $\iota=1$, $\ap=0$ and the left side vanishes identically: the marginal pledge
is fairly priced to the provider by the client, so it carries no signalling cost, and
\eqref{eq:newfoc} forces $D(s)=c_0+\chi/t$ wherever $\rho'>0$---a constant, incompatible with strict
continuous separation. With insurance the left side is nonzero and \eqref{eq:newfoc} is a linear ODE
whose solution is $D(s)=c_0+\chi/t+C[(\iota-1)(1-\rho(s))+\ap]^{1/(\iota-1)}$, but the participation
failure at the bottom remains. Numerically, on schedule \eqref{eq:oldmenu} with recovery in revenue,
a type $s=.8$ reports $\nappBrephi$ and a type $s=.1$ reports $\nappBreplo$.

The conclusion is not that the pledge does not separate. It is that it separates
\emph{coarsely}: a discrete menu deters a mimic at first order, through the jump from the null to a
positive pledge, while a continuous schedule asks each type to be deterred from its neighbour at
second order, which the fairly priced marginal pledge cannot deliver. Theorem~\ref{thm:binary} and
Proposition~\ref{prop:coarse} are the equilibrium objects that survive.

\clearpage
\section*{Declarations}
\noindent\emph{Competing interests.} The author is the author of \emph{Diebstahlsicher} (Vienna,
September 2026, ISBN 978-3-9506352-0-1), a German-language practitioner book that draws on this
line of work.
\emph{Funding.} None.
\emph{Data availability.} Replication code, parameters and random seeds for every figure and every
reported number are archived at doi:10.6084/m9.figshare.33193704.
\emph{Ethics.} No human subjects or proprietary data are involved.

\section*{Use of generative AI}
\noindent In preparing this manuscript the author used Anthropic's Claude, including Claude Code
(accessed July to September 2026), for four purposes: supplementary literature search and source
discovery; drafting and language editing of the text; implementing and running the numerical and
simulation code underlying the model calculations and the figures; and assistance with the
derivations. The research question, the model and its assumptions, the interpretation of the results
and all final decisions about content are the author's. AI technology was not used to generate, alter
or fabricate research data, and is not credited as an author. No AI-assisted output entered the
manuscript unverified: every reference was checked against the publisher record, and every reported
number is regenerated from the deposited replication code rather than transcribed. The author takes
full responsibility for the content of the manuscript, including all claims, citations and code.

\clearpage
\bibliography{refs}
\end{document}